\documentclass[conf]{new-aiaa}
\usepackage[utf8]{inputenc}

\usepackage{graphicx}
\usepackage{amsmath}
\usepackage{amsfonts}
\usepackage{gensymb}
\usepackage[version=4]{mhchem}
\usepackage{siunitx}
\usepackage{longtable,tabularx}
\usepackage{placeins}
\usepackage{xcolor}
\usepackage{multirow}

\newcommand{\MC}{\mathcal{C}}
\newcommand{\MJ}{\mathcal{J}}
\newcommand{\ML}{\mathcal{L}}

\newcommand{\MP}{\mathbb{P}}

\newcommand{\ith}{i^{\text{th}}}
\newcommand{\sgn}{\text{sgn}}
\newcommand{\Var}{\mathbf{Var}}

\title{
Aerodynamic Risk Assessment using Parametric, Three-Dimensional Unstructured, High-Fidelity CFD and Adaptive Sampling}
\author{Runda Ji \footnote{CFD Application Scientist.}}
\affil{Flexcompute Inc, Belmont, Massachusetts, 02138}
\author{Qiqi Wang \footnote{Associate Professor, Department of Aeronautics and Astronautics, AIAA Associate Fellow.}}
\affil{Massachusetts Institute of Technology, Cambridge, Massachusetts 02139}

\begin{document}

\maketitle

\begin{abstract}
We demonstrate an adaptive sampling approach for computing the probability of a rare event for a set of three-dimensional airplane geometries under various flight conditions.
We develop a fully automated method to generate parameterized airplanes geometries and create volumetric mesh for viscous CFD solution.
With the automatic geometry and meshing, we perform the adaptive sampling procedure to compute the probability of the rare event.
We show that the computational cost of our adaptive sampling approach is hundreds of times lower than a brute-force Monte Carlo method.
\end{abstract}

\section*{Nomenclature}

{\renewcommand\arraystretch{1.0}
\noindent\begin{longtable*}{@{}l @{\quad=\quad} l@{}}
$C_L$ & lift coefficient \\
$\MC$ & critical value of objective function \\
$\MJ$ & objective function \\
$\tilde{\MJ}$ & linear approximation of objective function \\
$M$ & Mach number \\
$N$ & desired number of sampling points in an adaptive iteration \\
$N_A$ & number of linear approximations \\
$N_i$ & number of sampling points in the $\ith$ stratum \\
$N_S$ & number of strata \\
$\MP$ & probability \\
$S_i$ & the $\ith$ stratum \\
$\Var$ & variance \\
$w$ & collection of all stochastic parameters \\
$\alpha$ & angle of attack \\
$\beta$ & side-slip angle \\
$\ML$ & Lagrangian function \\
$\sigma$ & root mean square deviation \\
$\lambda$ & Lagrange multiplier \\
\end{longtable*}}

\newpage

\section{Introduction}
Numerical simulations have been widely applied in aerodynamic shape design and optimization for decades.
Such numerical tools provide us more underlying physics with higher resolution, and most importantly, they are more affordable compared to wind tunnel tests.
However, the key restriction for such simulations is the computational time, drastically limiting the number of configurations/designs could be tested and analyzed within the required time-limit.
The entire workflow was so time-consuming that the users have to use naive Monte Carlo method because they did not have the capability/time to run multiple sampling iterations.
Unfortunately, if the total number of sampling points is limited, naive Monte Carlo method in high dimensional sampling space cannot allocate adequate sampling points within a specific region, and hence when estimating high-impact low-probability events, high stochastic error is almost inevitable for naive Monte Carlo method.


\bigskip

The idea of allocating sampling points adaptively was originated while characterizing the operability limits of a supersonic combustion engine to determine the safe operation region. \cite{iaccarino2011qmu}
Extra sampling points were allocated in the uncertain region between the safe operation and unstart of the engine. \cite{wang2012scramjet}
When performing an optimization under uncertainty, introducing a surrogate model can greatly reduce the computational cost, but the results are less accurate than the original model. \cite{eldred2002formulations, giunta2004perspectives}
Rather than using the approximations from surrogate models to directly replace real simulations, such approximations can be used to guide the allocation of real simulations.
Following this idea, we present an adaptive sampling procedure to accurately estimate the probability of a rare event in this article.

\bigskip

Specifically, we introduce three stochastic parameters in our parameterized transport airliner geometry: aspect ratio, sweep and dihedral angle of the wings.
Meanwhile, there are three freestream parameters: angle of attack, side-slip angle and the Mach number.
The parameterized geometries are first generated via Engineering Sketch Pad (ESP) \cite{haimes2013esp, dannenhoffer2016esp, haimes2017esp}, then meshed through Pointwise, \cite{Pointwise} and
finally the flow solutions are computed using Flow360.
In this paper, we focus on estimating the probability of exceeding a critical lift coefficient, as an example of estimating the probability of a rare event.
Once the lift coefficients are calculated, we build a linear regression to model the relationship between the lift coefficient and the geometric/freestream parameters.
This regression model is further applied to divide the stochastic parameters into multiple strata, while the optimal number of additional cases in each stratum is determined by two factors:
First, the probability a sample lies in this stratum.
Second, the conditional probability the rare event happens given the sample lies in this stratum.
After determining the optimal distribution of additional cases, we simulate additional 99 cases. 
Combining the lift coefficients obtained from 100 preliminary and additional 99 solutions, we efficiently and accurately estimate the probability of achieving a high-lift coefficient under stochastic geometric and freestream parameters.

\bigskip

For clarity, we summarize the adaptive sampling procedure as follows:
\begin{enumerate}
\item simulate 100 preliminary cases and calculate the real objective functions (lift coefficients) $\MJ(w) = C_L$
\item construct linear regression model based on the preliminary flow solutions
\item generate a lot (10,000,000) of stochastic parameters $w$, and calculate the linearly approximated objective function $\tilde{\MJ}(w)$
\item define the strata based on the linear approximation $\tilde{\MJ}(w)$
\item estimate the probability $P_i^{(1)}$ that a sample lies in the $\ith$ stratum $S_i$
\item estimate the conditional probability $P_i^{(2)}$ that the real objective function $\MJ(w) > 0.9$ given the sample lies in the $\ith$ stratum $S_i$
\item determine the optimal number of additional cases $N_i$ in each stratum
\item run additional cases according to the distribution of $N_i$
\item re-calculate the conditional probability $P_i^{(2)}$ based on the 100 preliminary plus the 99 additional flow solutions
\item estimate $\MP(\MJ(w) > 0.9)$ and its variance based on $P_i^{(1)}$ and $P_i^{(2)}$
\end{enumerate}

\newpage

\section{Preparation}
Before initiating the sampling process, there are several prerequisites.
First, we need to automatically generate a considerable number of parameterized geometry files.
Secondly, we also need to mesh these geometry files in batch mode.
Finally, we need the capability to automatically launch and post-process the cases.

\subsection{Automatic Generation of Parameterized Geometries}
In this paper, we use the Engineering Sketch Pad (ESP) to generate parameterized geometries automatically.
The ranges of geometric parameters are listed as follows,
\begin{table}[h!]
\centering
\begin{tabular}{c c c c}
\hline
\hline
Parameter & Min & Original & Max \\
\hline
Aspect ratio      & 5.0  & $\approx10.1$ & 15.0 \\
Sweep ($\deg$)    & 25.0 & 35.0 & 45.0 \\
Dihedral ($\deg$) & -5.0 &  4.0 & 15.0 \\
\hline
\hline
\end{tabular}
\caption{Range of geometric parameters.}
\label{tab_geom_parms}
\end{table}

For clarity, the parameterized geometries with various aspect ratio, sweep and dihedral angle are shown as follows.
\begin{figure}[h!]
  \centering
  \includegraphics[height=6cm,trim=25cm 0cm 40cm 0cm,clip]{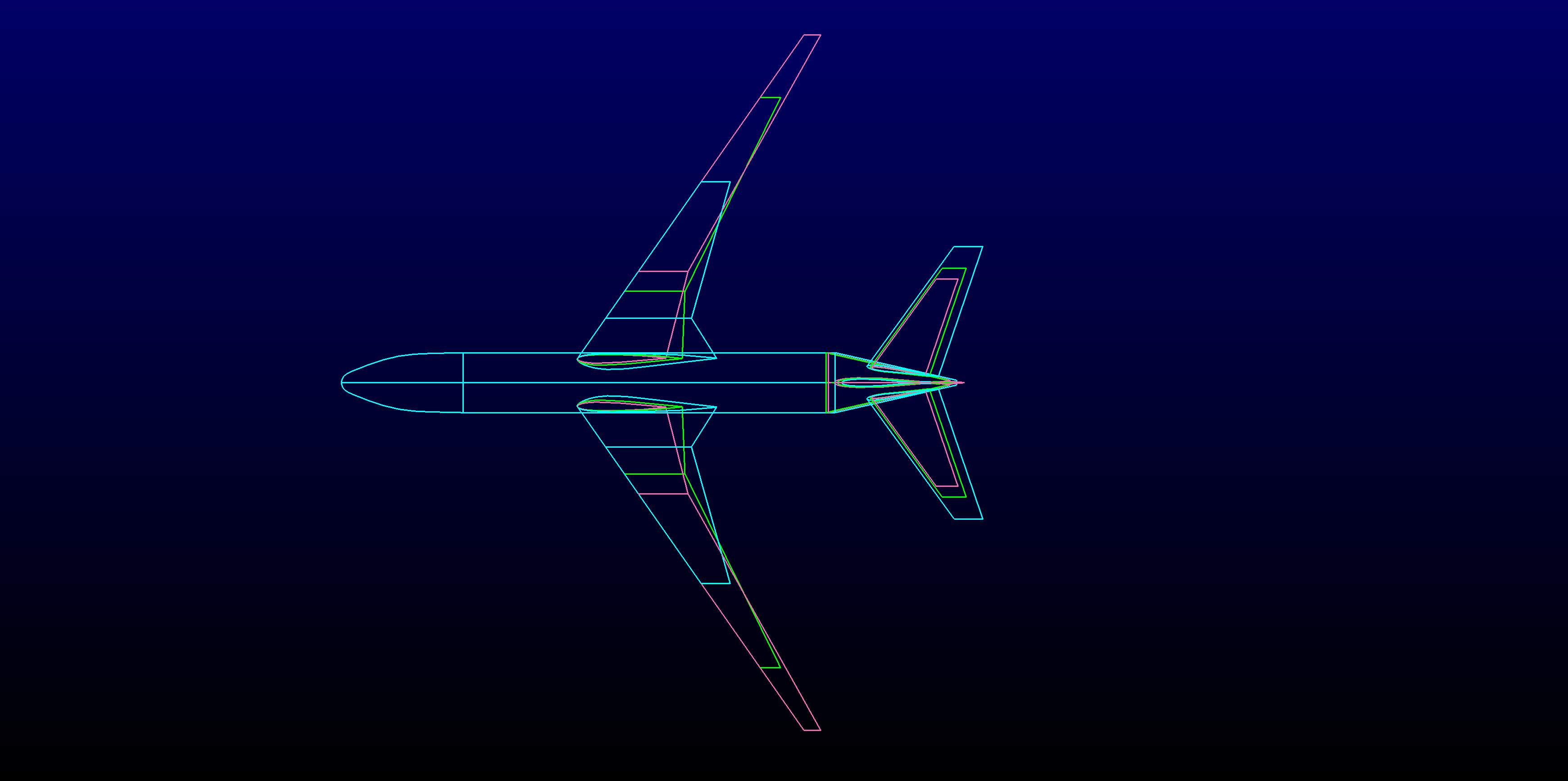}
  \includegraphics[height=6cm,trim=25cm 0cm 30cm 0cm,clip]{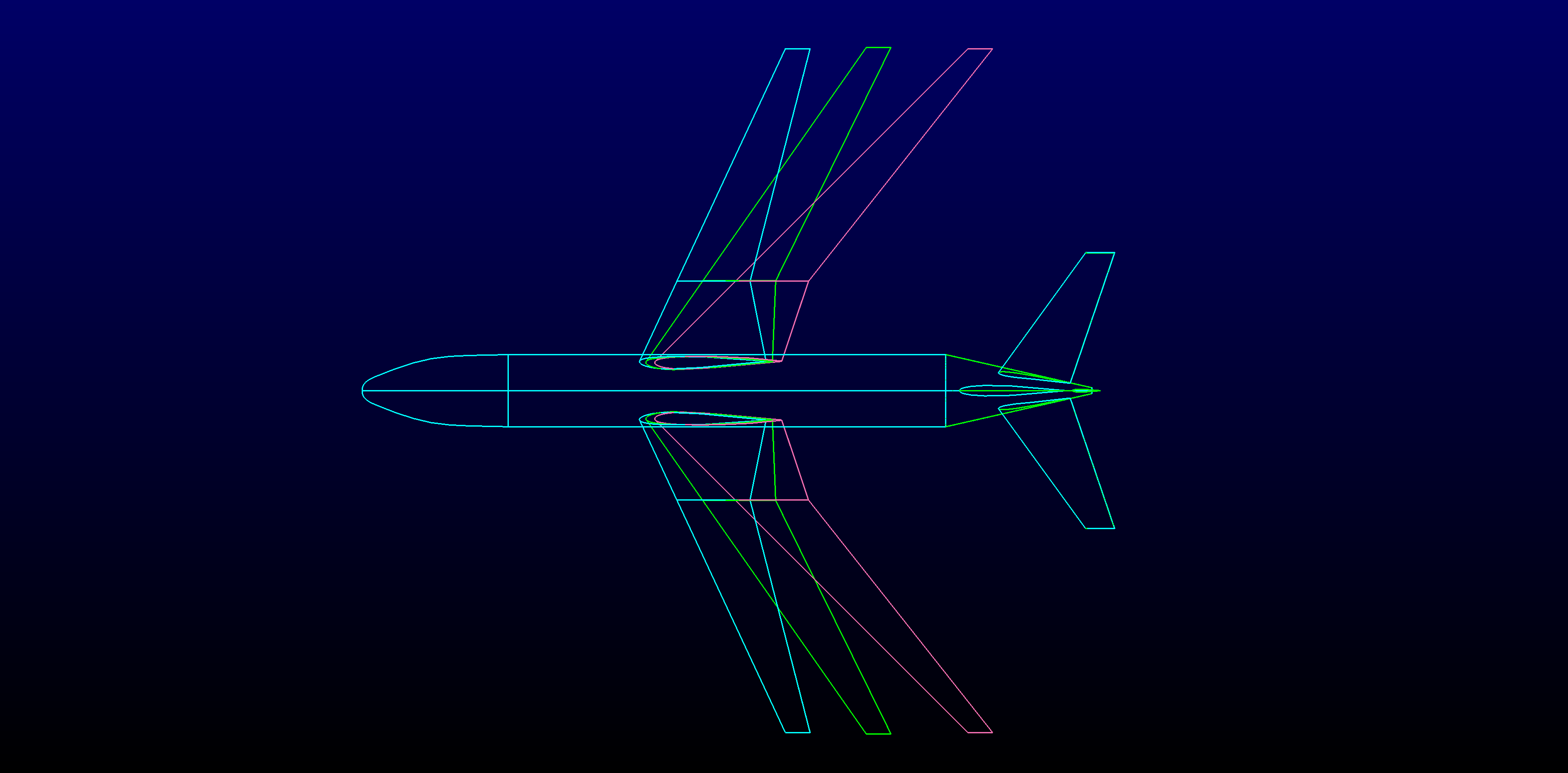}
  \includegraphics[height=6cm,trim=15cm 0cm 30cm 0cm,clip]{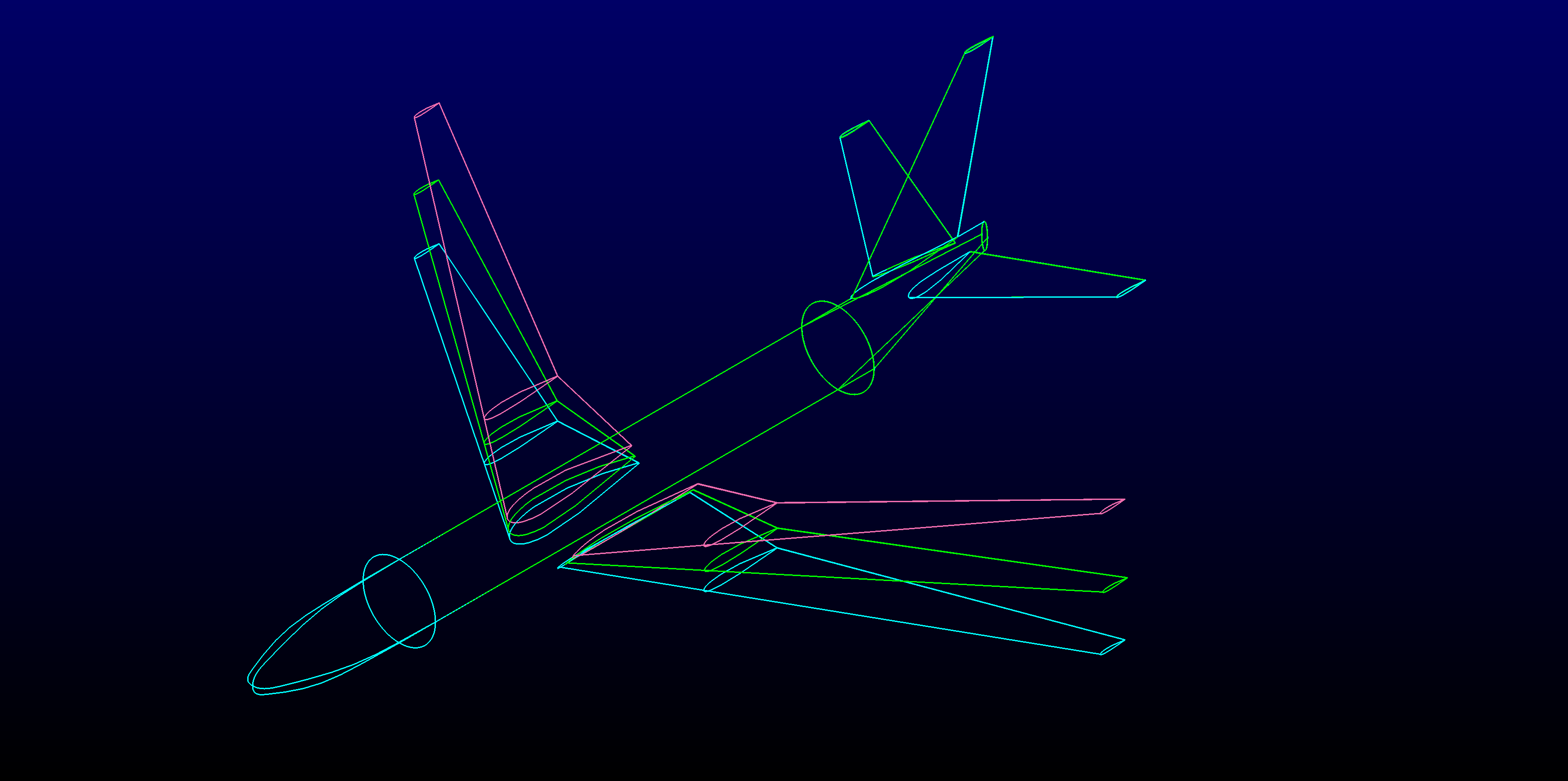}
  \caption{Parameterized geometry: variation of aspect ratio, sweep and dihedral angle.}
  \label{fig_param_geom}
\end{figure}
\FloatBarrier

\subsection{Automatic Generation of Mesh}
After the parameterized geometries are generated, instead of meshing each geometry manually, we use Pointwise Glyph to script the entire mesh generation process.
The surface and volume meshes obtained from this fully-automatic meshing process are shown as follows.
\begin{figure}[h!]
  \centering
  \includegraphics[height=6cm,trim=20cm 0cm 20cm 0cm,clip]{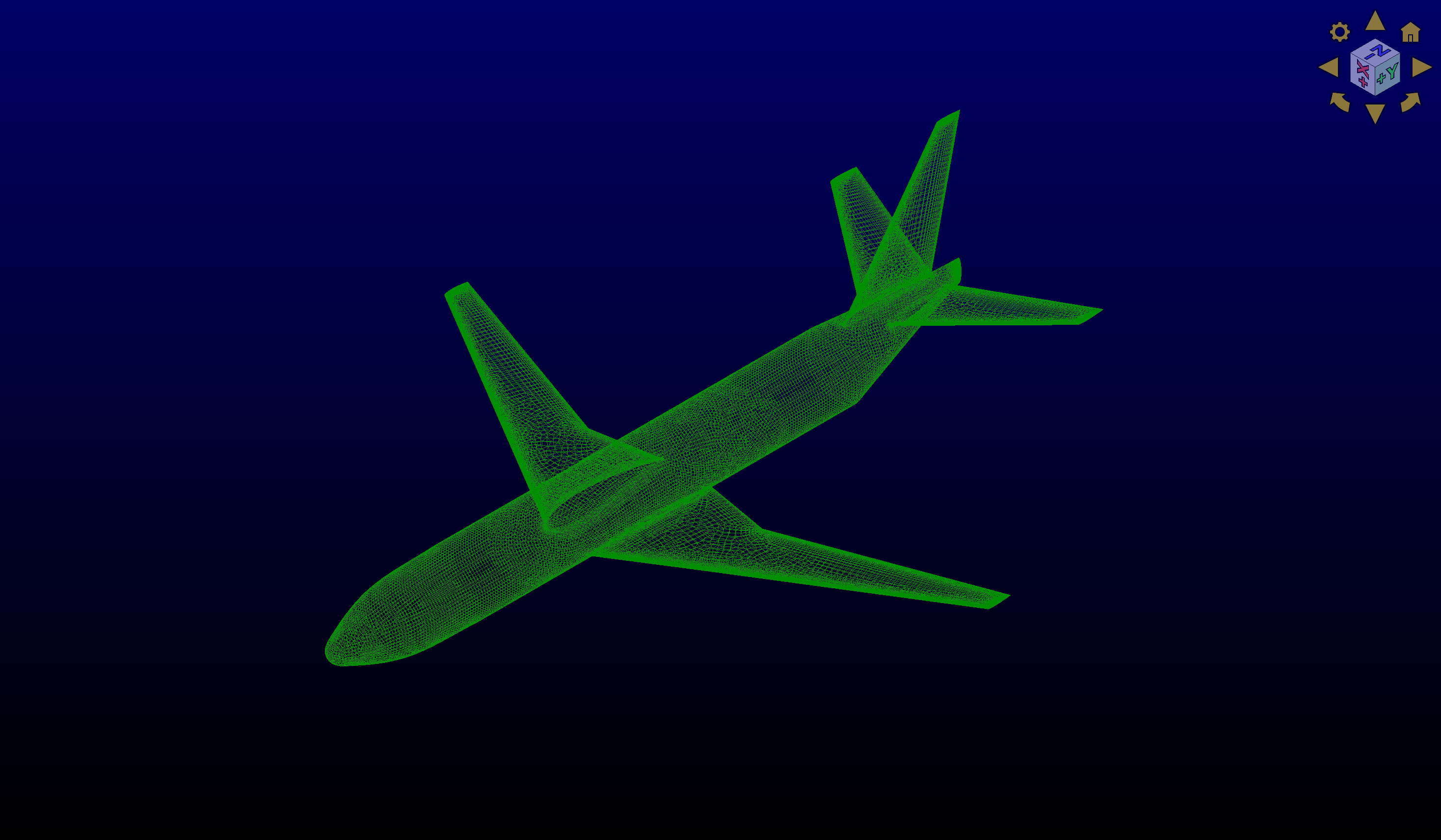}
  \includegraphics[height=6cm,trim=5cm 0cm 15cm 0cm,clip]{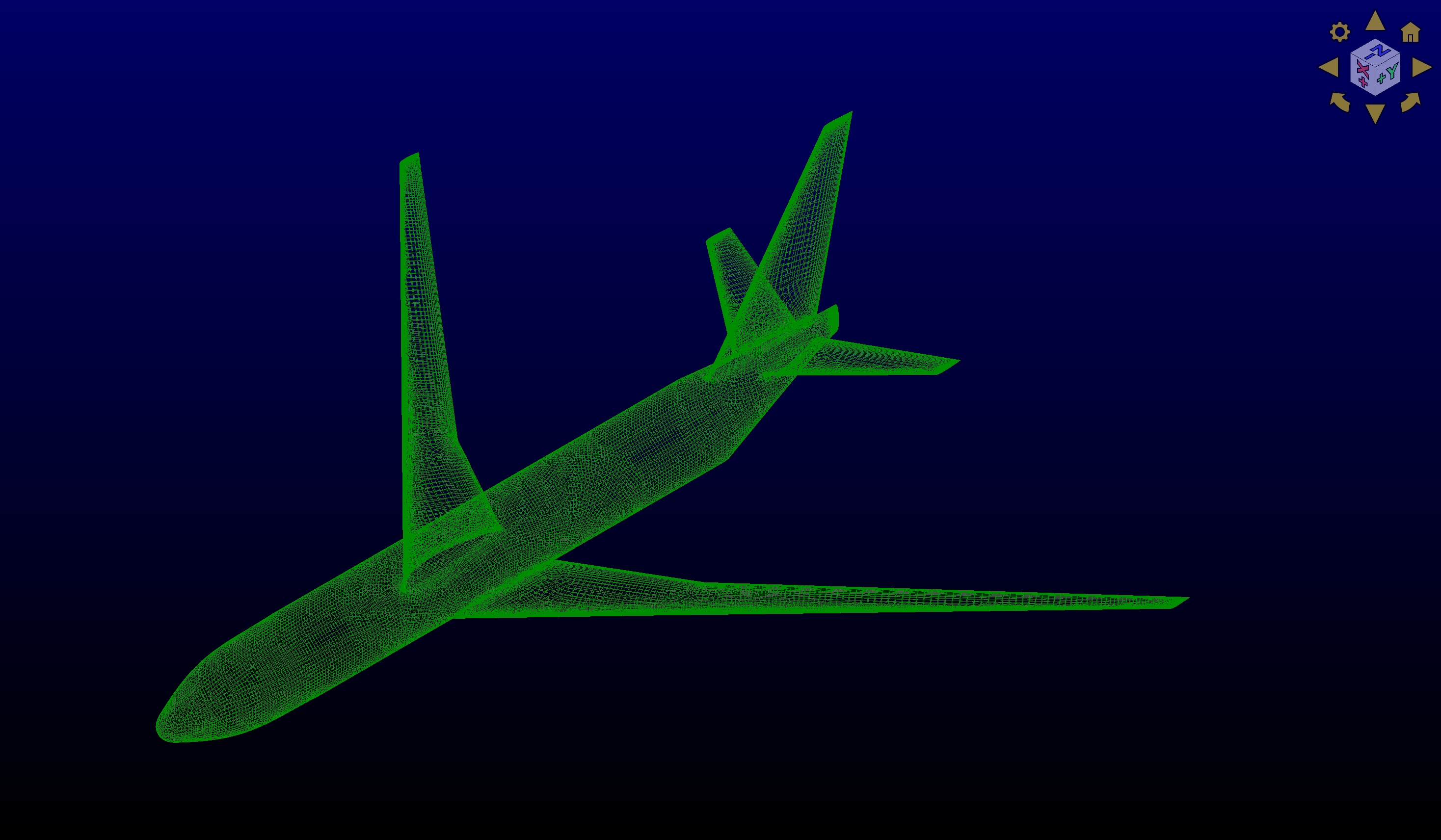}
  \caption{Surface meshes for parameterized geometries automatically generated using Pointwise Glyph script.
  Left:  aspect ratio $\approx 8.53$, sweep $\approx 30.51\degree$, dihedral $\approx 0.94\degree$.
  Right: aspect ratio $\approx 14.95$, sweep $\approx 44.71\degree$, dihedral $\approx 1.47\degree$.}
  \label{fig_surf_mesh}
\end{figure}
\FloatBarrier

\begin{figure}[h!]
  \centering
  \includegraphics[height=7cm,trim=20cm 0cm 20cm 0cm,clip]{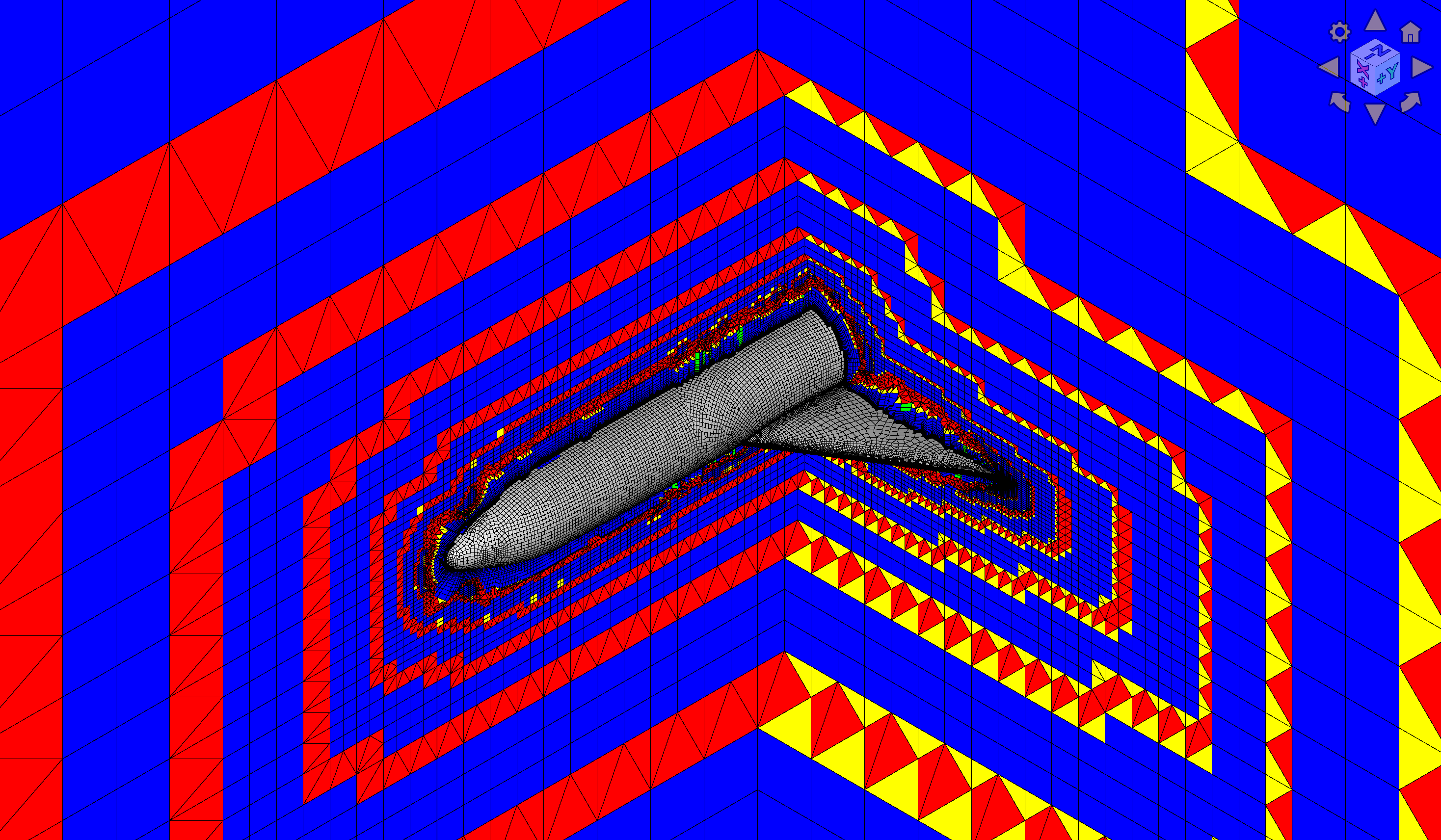}
  \includegraphics[height=7cm,trim=15cm 0cm 20cm 0cm,clip]{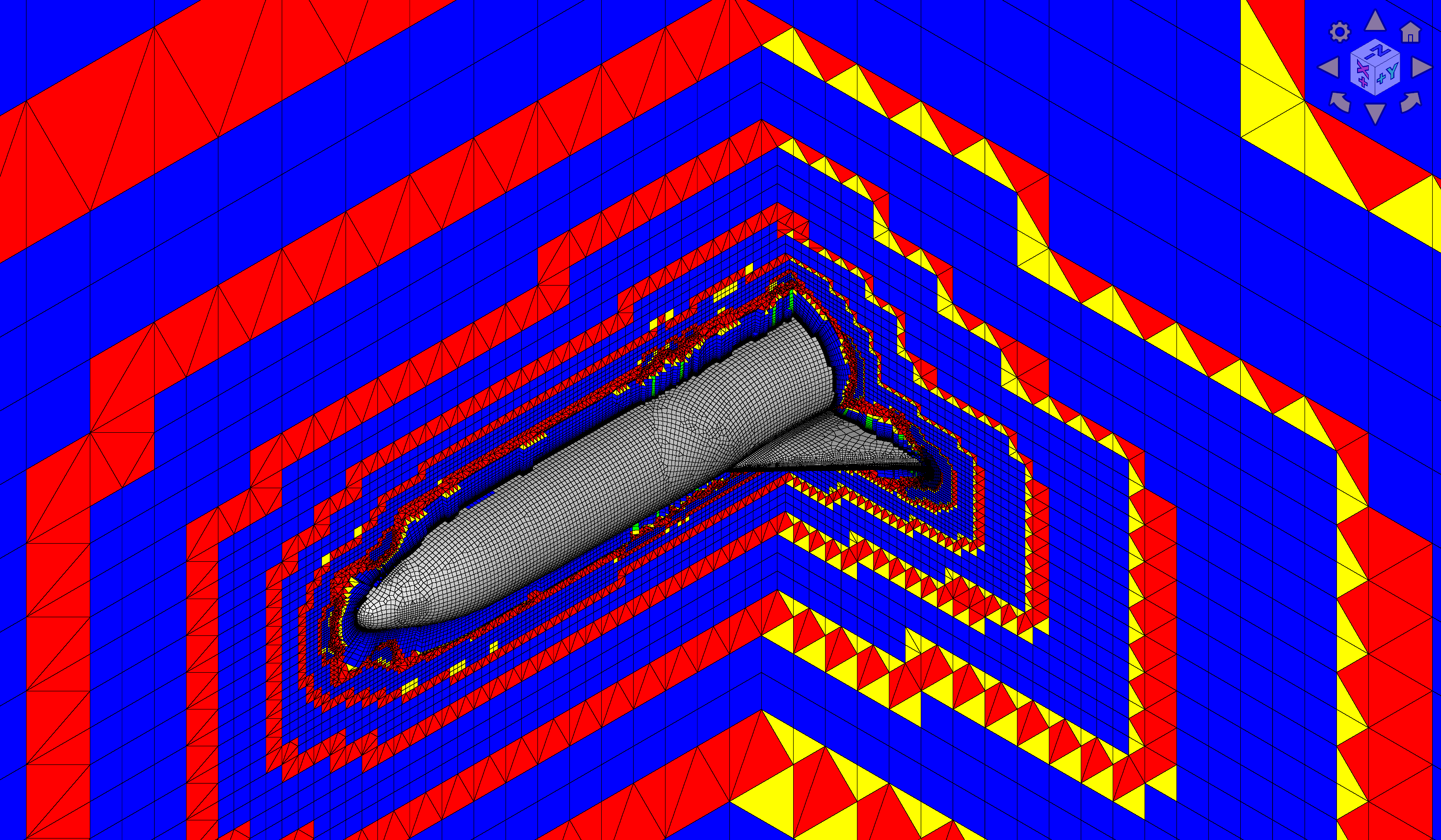}
  \caption{Volume meshes for parameterized geometries automatically generated using Pointwise Glyph script. The cells are colored by different types. Red: Tets. Yellow: Pyramids. Green: Prisms. Blue: Hexes.}
  \label{fig_volume_mesh}
\end{figure}
\FloatBarrier

\subsection{CFD Simulations}
After obtaining these mesh files, we now want to run simulations under different freestream conditions.
Rather than setting up the solver manually for each case, we create a template for generating solver configuration files.
By substituting the freestream parameters into this template, the configuration file for each case is automatically generated.
Utilizing the Flow360 PythonAPI, we can easily upload the mesh and configuration files to the cluster, and the simulation will start automatically.

\newpage

For clarity, the ranges of freestream parameters are listed as follows,
\begin{table}[ht!]
\centering
\begin{tabular}{c c c}
\hline
\hline
Parameter & Min & Max \\
\hline
Angle of attack $\alpha$  & 0.0 & 8.0 \\
Side-slip angle $\beta$   & 0.0 & 5.0 \\
Mach number               & 0.1 & 0.3 \\
\hline
\hline
\end{tabular}
\caption{Range of freestream parameters.}
\label{tab_freestream_parms}
\end{table}
\FloatBarrier

For demonstration, the numerical results under different freestream conditions are shown as follows.
The airplanes displayed here are the exactly the same with the airplanes shown in previous section.
\begin{figure}[h!]
  \centering
  \includegraphics[width=8cm,trim=2cm 0cm 2cm 0cm,clip]{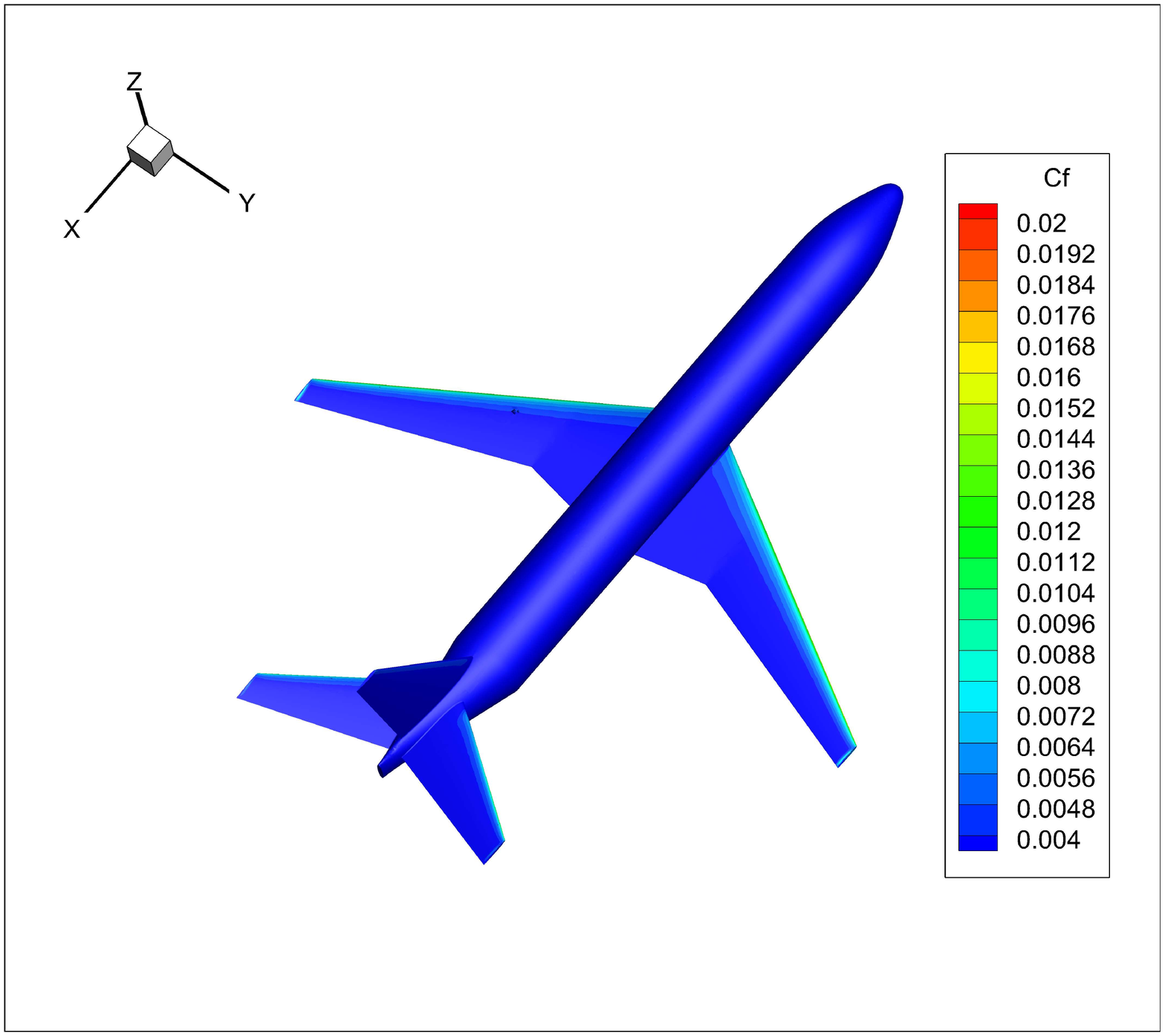}
  \includegraphics[width=8cm,trim=2cm 0cm 2cm 0cm,clip]{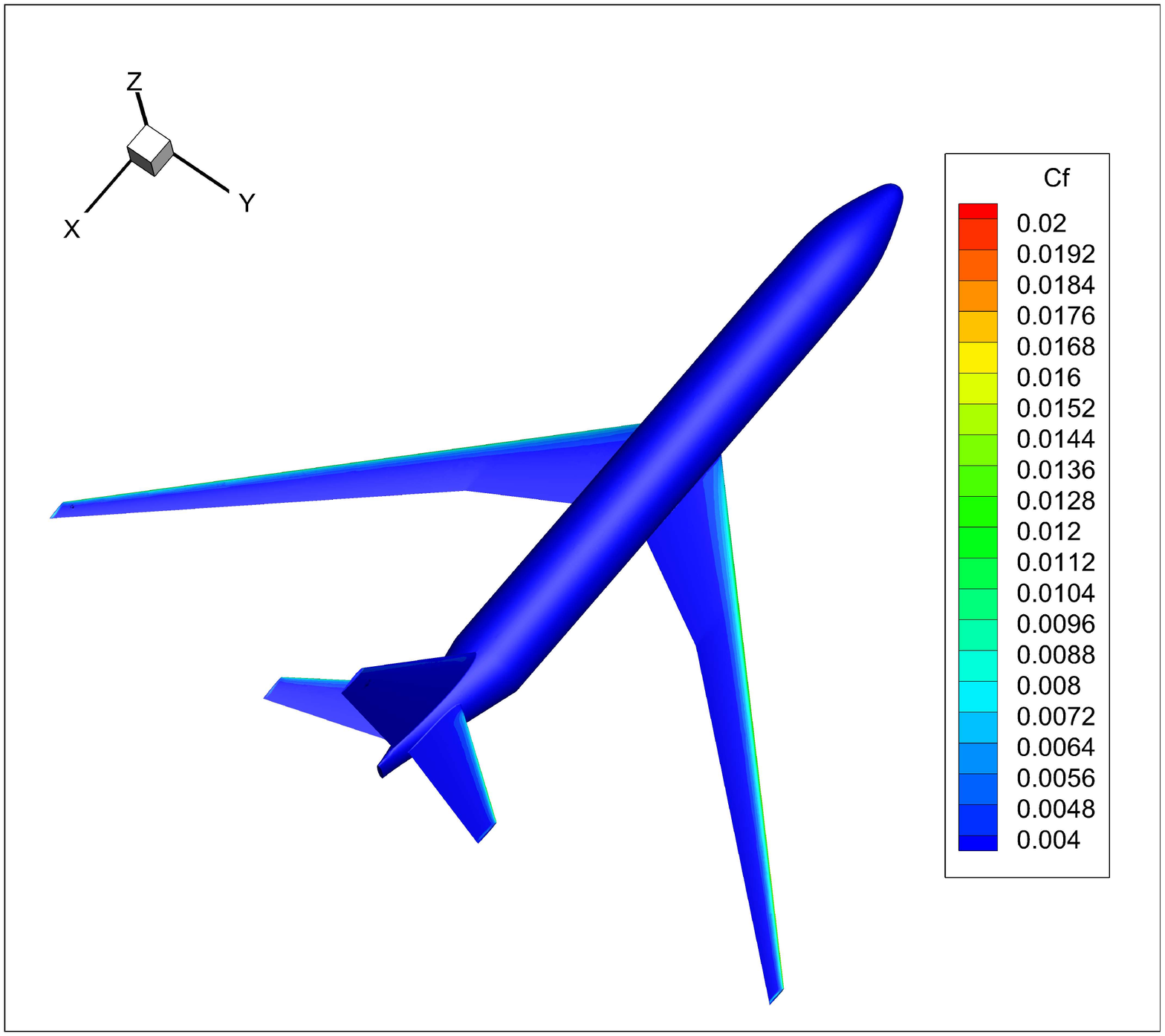}
  \caption{Friction coefficient $C_f$ distribution on parameterized geometries under different freestream conditions.
  Left:  $\alpha \approx 7.59\degree$, $\beta \approx 1.12\degree$, Mach$\approx 0.21$.
  Right: $\alpha \approx 7.68\degree$, $\beta \approx 1.76\degree$, Mach$\approx 0.18$.}
  \label{fig_Cf}
\end{figure}
\FloatBarrier

\begin{figure}[h!]
  \centering
  \includegraphics[width=8cm,trim=1cm 0cm 1cm 0cm,clip]{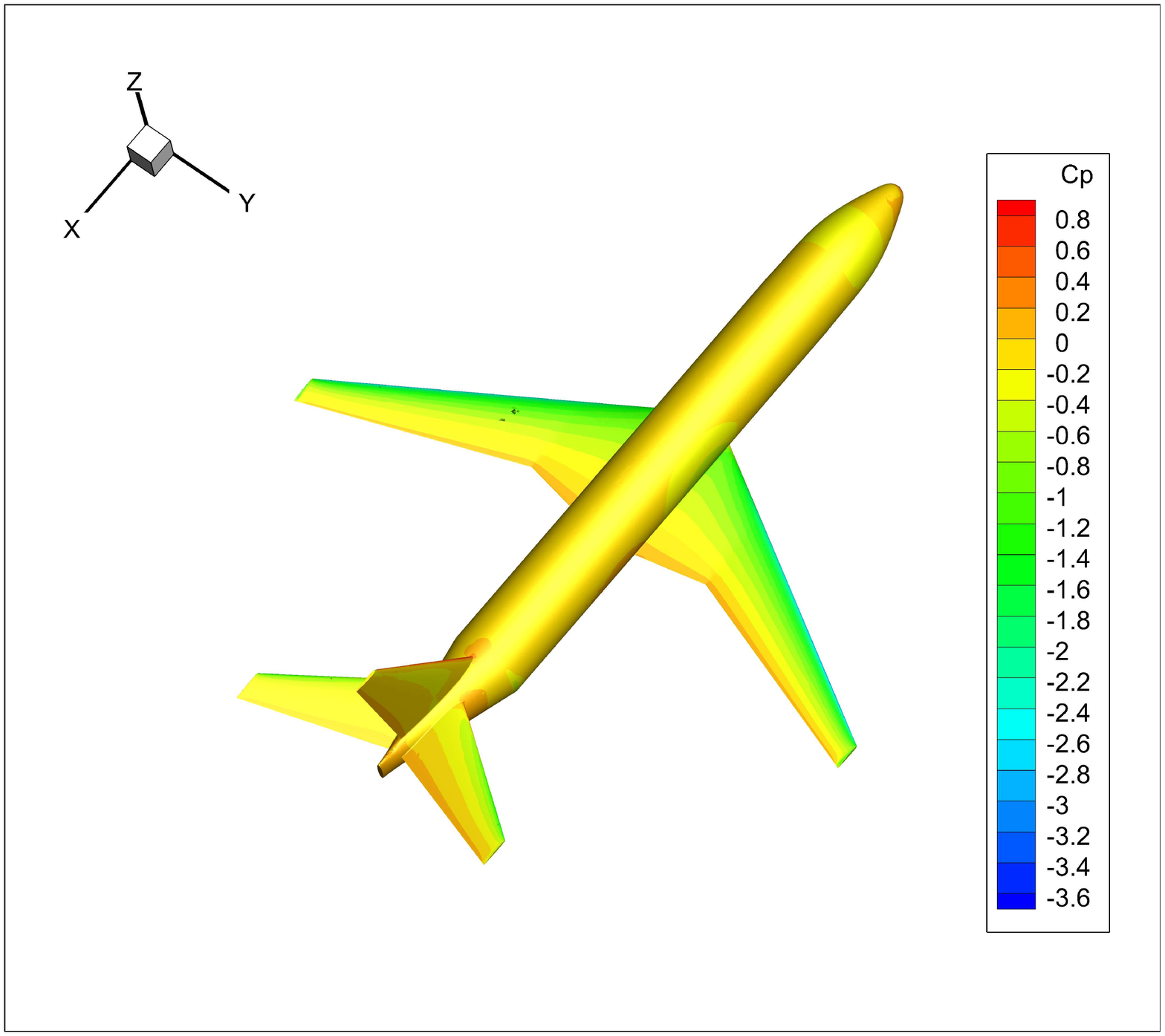}
  \includegraphics[width=8cm,trim=1cm 0cm 1cm 0cm,clip]{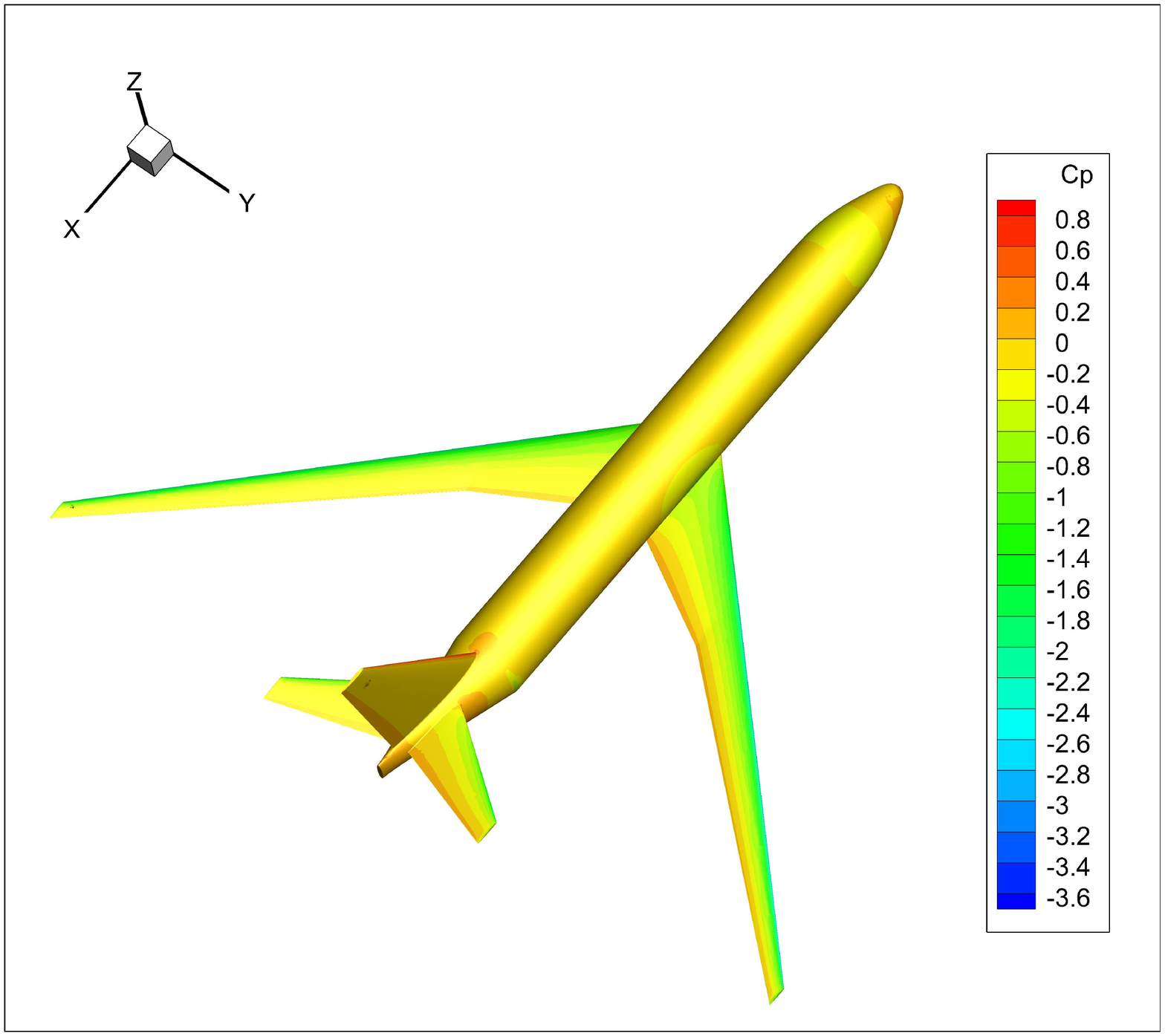}
  \caption{Pressure coefficient $C_p$ distribution on parameterized geometries under different freestream conditions.
  Left:  $\alpha \approx 7.59\degree$, $\beta \approx 1.12\degree$, Mach$\approx 0.21$.
  Right: $\alpha \approx 7.68\degree$, $\beta \approx 1.76\degree$, Mach$\approx 0.18$.}
  \label{fig_Cp}
\end{figure}
\FloatBarrier

\begin{figure}[h!]
  \centering
  \includegraphics[width=8cm,trim=1cm 0cm 1cm 0cm,clip]{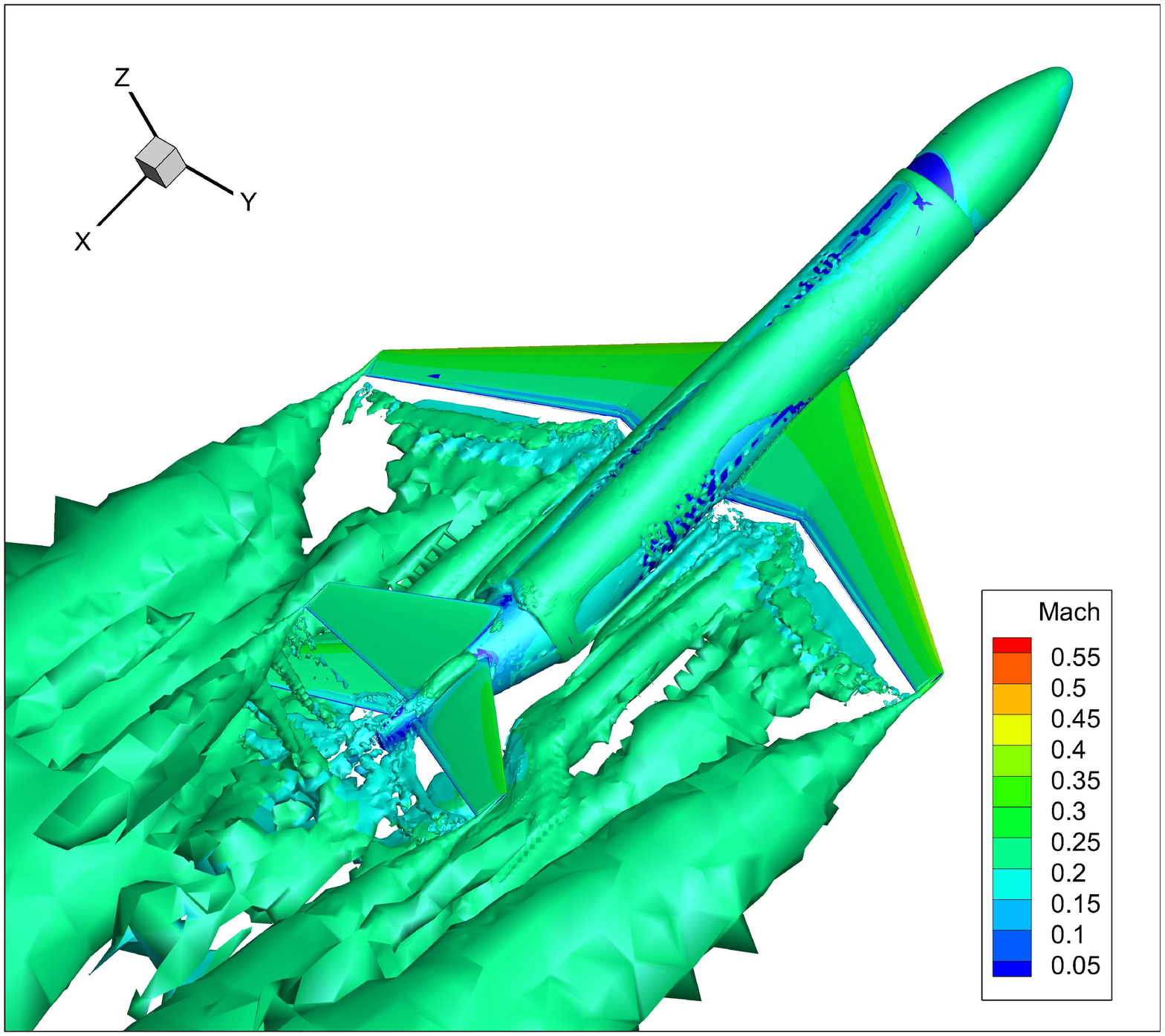}
  \includegraphics[width=8cm,trim=1cm 0cm 1cm 0cm,clip]{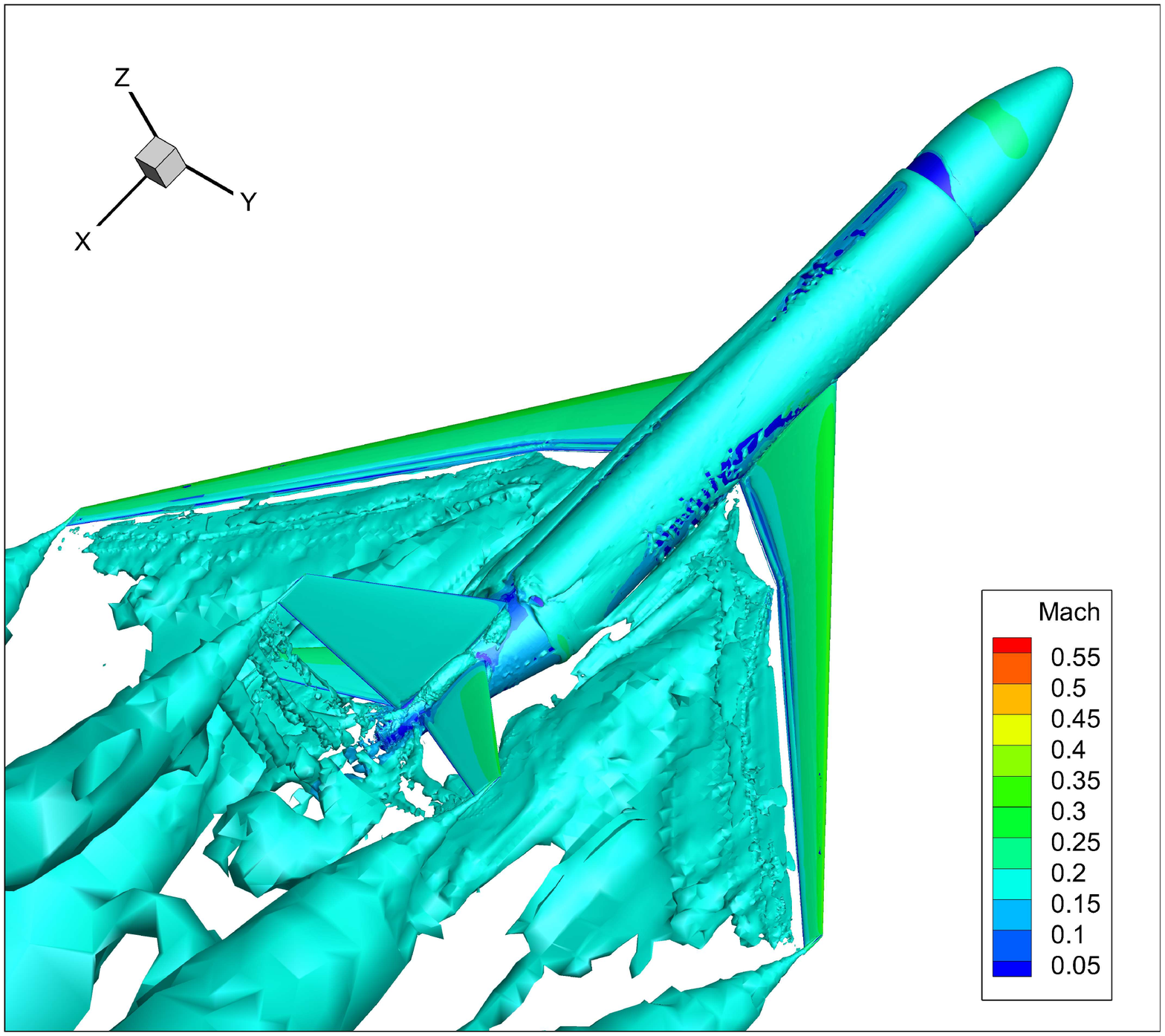}
  \caption{Q-criterion for parameterized geometries under different freestream conditions.}
  \label{fig_Q_criterion}
\end{figure}
\FloatBarrier

\section{Optimizing the Allocation of Samples: Theory}
In this section, we will elaborate the theory of adaptive sampling approach and further derive the optimized distribution for sampling points.

\subsection{Linear Regression of the Preliminary Flow Solutions}
Since we want to accurately estimate the probability of $C_L > 0.9$, we set the lift coefficient as our objective function $\MJ(w) = C_L$.
Once the objective function is determined, we perform a linear regression between the objective function $\MJ(w)$ and the stochastic parameter vector
$w = (\text{aspect ratio}, \text{sweep}, \text{dihedral}, \alpha, \beta, \text{Mach})$.

\bigskip

For the 100 preliminary cases, the parameters are evenly distributed in the sampling space.
We first generate 10 geometries, and then for each geometry we run 10 simulations under 10 different freestream conditions.
The real $\MJ(w) = C_L$ obtained from the flow simulations versus the linearly approximated $\tilde{\MJ}(w)$ is plotted as follows:

\begin{figure}[h!]
  \centering
  \includegraphics[height=7cm,trim=0cm 0.5cm 0cm 1.5cm,clip]{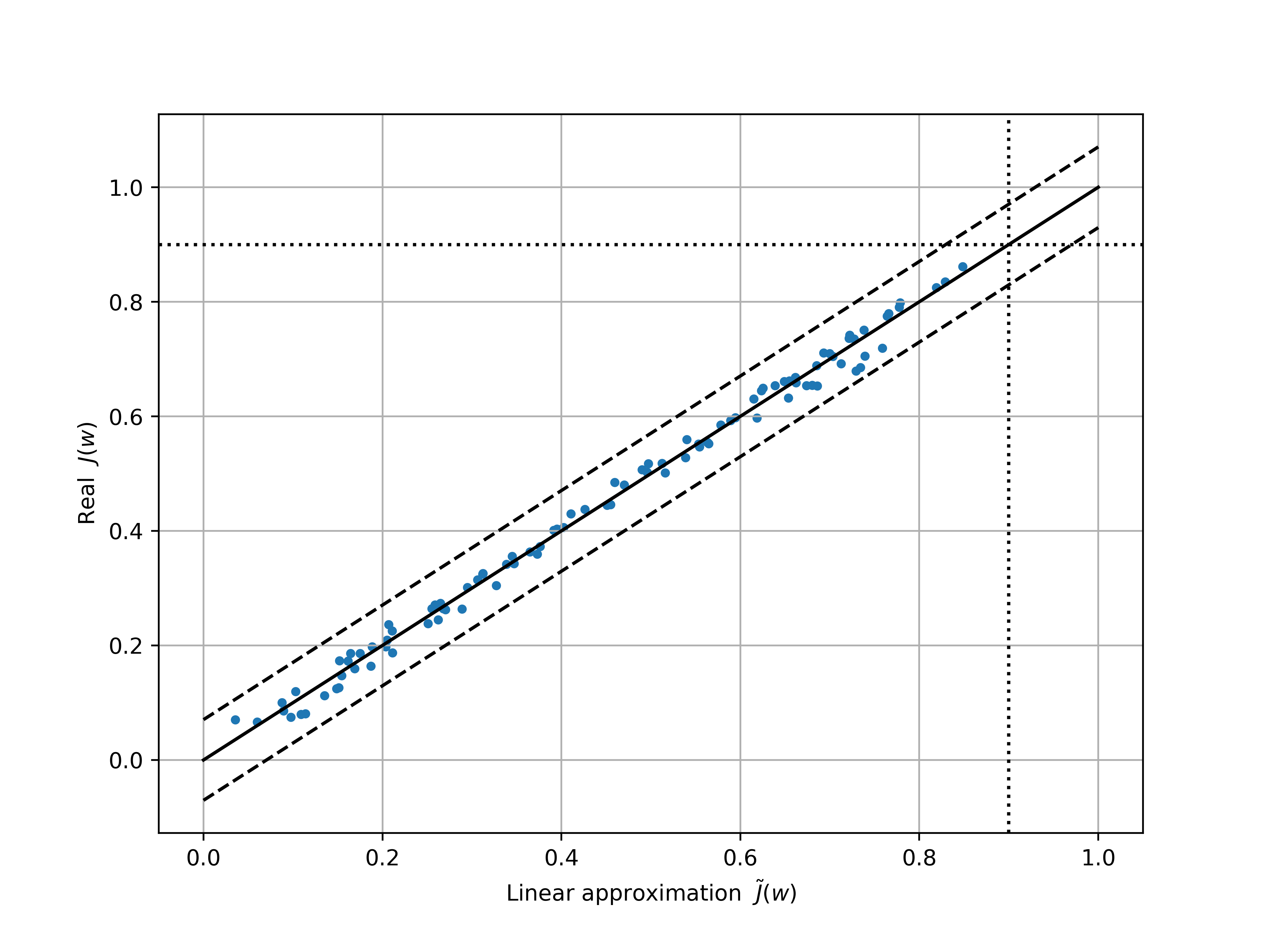}
  \caption{Real objective function $\MJ(w)$ versus the linear approximation $\tilde{\MJ}(w)$. The black solid line indicates the exact approximation, the dash line indicates the $\pm 4 \sigma$ range, where $\sigma$ is the root mean square deviation of the linear regression. The horizontal and vertical dotted lines show $\MJ(w) = 0.9$ and $\tilde{\MJ}(w) = 0.9$, respectively.}
  \label{fig_real_vs_approx_iter0}
\end{figure}
\FloatBarrier

We are interested in estimating $\MP(\MJ(w) > 0.9)$, so we set the critical lift coefficient $\MJ_{\text{critical}} = \MC = 0.9$.
Note that we have $\MJ(w) < 0.9$ for all 100 preliminary cases, which indicates the rare event does not happen in these preliminary cases, and hence we need to run additional simulations clustering around $\MJ(w) = 0.9$.

\subsection{Definition of Strata}
In this paper, we use "strata" to mathematically describe the allocation of sampling points.
Each stratum is a set of stochastic parameters $w$, such that the corresponding linear approximations $\tilde{\MJ}(w)$ fall in a certain range.
\begin{equation}
S_i = \{w | a < \tilde{\MJ}(w) < b \}
\end{equation}

For example, a possible way to define 1,000 strata in $(0,1)$ is
\begin{equation}\label{eqn_example_strata}
\begin{split}
S_1 &= \{w | 0 < \tilde{\MJ}(w) < 0.001 \} \\
S_2 &= \{w | 0.001 < \tilde{\MJ}(w) < 0.002 \} \\
S_3 &= \{w | 0.002 < \tilde{\MJ}(w) < 0.003 \} \\
&\cdots
\end{split}
\end{equation}

Note that the construction of strata is arbitrary and we need to adjust the size and location of the strata based on our problem, i.e. we are not using equation (\ref{eqn_example_strata}) to define the strata in the following discussions.

\subsection{The Minimization Problem}
The goal of accurately estimating the probability of achieving a high lift coefficient can be converted in to a minimization problem,
\begin{equation}
\min_{N_i} \bigg\{ \text{Var} \big[\MP(\MJ(w) > \MC)\big] \bigg\}
\end{equation}

where $N_i$ is the number sampling points within the $\ith$ stratum.
$\MP(\MJ(w) > \MC)$ can be re-written as,
\begin{equation}
\MP(\MJ(w) > \MC) = \sum_{i=1}^{N_S} \MP (w \in S_i) \MP (\MJ(w) > \MC | w \in S_i)
\end{equation}
where $\MP (w \in S_i)$ is the probability of a sample lies in the $\ith$ stratum,
while $\MP (\MJ(w) > \MC | w \in S_i)$ is the conditional probability that the real objective function $\MJ(w) > \MC$ under the condition that $w \in S_i$,
and $N_S$ is the total number of strata.

\subsection{Probability of a sample lies in the $\ith$ stratum}
To begin with, let us consider the first term $\MP (w \in S_i)$, which can be estimated by,
\begin{equation}
\MP (w \in S_i) 
\approx P_i^{(1)} 
= \frac{1}{N_A} \sum_{k=1}^{N_A} I_{w_k \in S_i}
\end{equation}

where $N_A$ is the total number of linearly approximation $\tilde{\MJ}(w)$, in our project we set $N_A = 10,000,000$.
For the $k^{\text{th}}$ sample,
\begin{equation}
I_{w_k \in S_i} =
\begin{cases}
  1, & \mbox{if } w_k \in S_i \\
  0, & \mbox{if } w_k \notin S_i \\
\end{cases}
\end{equation}

The biased sample variance of $P_i^{(1)}$ can be written as,
\begin{equation}
\begin{split}
\Var\big[ P_i^{(1)} \big] 
&= \Var \bigg[ \frac{1}{N_A} \sum_{k=1}^{N_A} I_{w_k \in S_i} \bigg] \\
&= \bigg(\frac{1}{N_A}\bigg)^2 \sum_{k=1}^{N_A} \Var \big[ I_{w_k \in S_i} \big]
\end{split}
\end{equation}

where for each $w_k \in S_i$,
\begin{equation}
\Var\big[ I_{w_k \in S_i} \big] = P_i^{(1)} \big( 1-P_i^{(1)} \big)
\end{equation}

Hence we have,
\begin{equation}
\Var \big[ P_i^{(1)} \big] 
= \frac{P_i^{(1)} \big( 1-P_i^{(1)} \big)}{N_A}
\end{equation}

Meanwhile the unbiased population variance can be written as,
\begin{equation}
\begin{split}
\big(s_i^{(1)}\big)^2 
&= \frac{N_A}{N_A-1} \frac{P_i^{(1)} \big( 1-P_i^{(1)} \big)}{N_A} \\
&= \frac{P_i^{(1)} \big( 1-P_i^{(1)} \big)}{N_A - 1} \\
\end{split}
\end{equation}

When the total number of linear approximation $N_A$ is large, both $\Var \big[ P_i^{(1)} \big]$ and $\big(s_i^{(1)}\big)^2$ are negligible,
which indicates we may reasonably assume that $P_i^{(1)}$ is a constant rather than a random variable in further calculation.

\subsection{Conditional probability}
$\MP (\MJ(w) > \MC | w \in S_i)$ is the conditional probability that the real objective function $\MJ(w) > \MC$ under the condition $w \in S_i$,
\begin{equation}
\MP (\MJ(w) > \MC | w \in S_i) \approx
P_i^{(2)} = \frac{1}{N_i} \sum_{k=1}^{N_i}I_{\MJ(w_k^i)>\MC}
\end{equation}

As for the $k^{\text{th}}$ sample in the $\ith$ stratum $w_k^i$,
\begin{equation}
I_{\MJ(w_k^i)>\MC} =
\begin{cases}
  1, & \mbox{if } \MJ(w_k^i)>\MC \\
  0, & \mbox{if } \MJ(w_k^i)<\MC \\
\end{cases}
\end{equation}

The biased sample variance of $P_i^{(2)}$ can be written as,
\begin{equation}
\begin{split}
\Var \big[ P_i^{(2)} \big]
&= \Var\bigg[ \frac{1}{N_i} \sum_{k=1}^{N_i} I_{\MJ(w_k^i)>\MC} \bigg] \\
&= \frac{1}{N_i} \Var \big[ I_{\MJ(w_k^i)>\MC} \big] \\
&= \frac{P_i^{(2)} \big( 1 - P_i^{(2)}\big)}{N_i} \\
\end{split}
\end{equation}

Meanwhile the unbiased population variance can be written as,
\begin{equation}
\begin{split}
\big( s_i^{(2)} \big)^2 
&= \frac{N_i}{N_i-1} \frac{P_i^{(2)} \big( 1-P_i^{(2)} \big)}{N_i} \\
&= \frac{P_i^{(2)} \big( 1-P_i^{(2)} \big)}{N_i - 1} \\
\end{split}
\end{equation}

\subsection{The Optimized Distribution}
Assuming $P_i^{(1)}$ are constants, the biased sample variance of the estimated probability of the rare event can be written as,
\begin{equation}
\begin{split}
\Var \bigg[\sum_{i=1}^{N_S} P_i^{(1)} P_i^{(2)}\bigg]
&= \sum_{i=1}^{N_S} \big(P_i^{(1)}\big)^2  \Var \big[ P_i^{(2)}\big] \\
&= \sum_{i=1}^{N_S} \big(P_i^{(1)}\big)^2  \frac{P_i^{(2)} \big( 1 - P_i^{(2)}\big)}{N_i} \\
\end{split}
\end{equation}

In total, we have $N_S$ strata: $S_1, S_2, S_3 \cdots S_{N_S}$, denote the number of samples in the $\ith$ stratum $S_i$ as $N_i$
\begin{equation}
f(N_1, N_2, N_3 \cdots N_{N_S}) 
= \sum_{i=1}^{N_S} \big(P_i^{(1)}\big)^2 
  \frac{P_i^{(2)}\big(1 - P_i^{(2)}\big)}{N_i}
\end{equation}

The restriction is given by,
\begin{equation}
g(N_1, N_2, N_3 \cdots N_{N_S}) = \sum_{i=1}^{N_S} N_i - N = 0
\end{equation}
where $N$ is the total number of flow simulations we need to run.
The Lagrangian function can be constructed as,
\begin{equation}
\ML = f(N_1, N_2, N_3 \cdots N_{N_S}) + \lambda g(N_1, N_2, N_3 \cdots N_{N_S})
\end{equation}
where $\lambda$ is the Lagrange multiplier.

\begin{equation}
\begin{split}
\frac{\partial \ML}{\partial N_i} 
&= \big(P_i^{(1)}\big)^2 \frac{P_i^{(2)}\big(1 - P_i^{(2)}\big)}{-N_i^2} + \lambda \\
\frac{\partial \ML}{\partial \lambda} &= \sum_{i=1}^{N_S} N_i - N = 0
\end{split}
\end{equation}

which gives the number of sampling points $N_i$ in the $\ith$ stratum,
\begin{equation}
N_i \propto P_i^{(1)} \sqrt{P_i^{(2)}\big(1 - P_i^{(2)}\big)}
\end{equation}

Once we know $P_i^{(1)}$ and $P_i^{(2)}$, we can easily determine the optimal number of sampling points for each stratum.
Therefore, the key is to numerically estimate $P_i^{(1)}$ and $P_i^{(2)}$.

\section{Optimizing the Allocation of Samples: Implementation}
\subsection{Estimating the probability a sample lies in the $\ith$ stratum}
We generated $N_A = 10,000,000$ stochastic parameters and then compute the linear approximation $\tilde{\MJ}(w)$ for each $w$.
As aforementioned, the probability that a stochastic parameter vector $w$ lies in $S_i$ was estimated via
\begin{equation}
P_i^{(1)} = \frac{1}{N_A} \sum_{k=1}^{N_A} I_{w_k \in S_i}
\end{equation}

\subsection{Estimating the conditional probability}
$P_i^{(2)}$ is the conditional probability that the real objective function $\MJ(w) > \MC$ under the condition that $w \in S_i$
\begin{equation}
P_i^{(2)} = \frac{1}{N_i} \sum_{k=1}^{N_i}I_{\MJ(w_k^i)>\MC}
\end{equation}
Since each stratum $S_i$ is defined based on the linear approximation $\tilde{\MJ}(w)$, $P_i^{(2)}$ could be interpreted as an evaluation of our linear approximation.
Specifically, if our linear approximation were 100\% accurate, then we would have,
\begin{equation}
\begin{split}
\MP (\MJ(w) > \MC | \tilde{\MJ}(w) < \MC) &= 0 \\
\MP (\MJ(w) > \MC | \tilde{\MJ}(w) > \MC) &= 1 \\
\end{split}
\end{equation}

Although our linear regression cannot be 100\% accurate, it does offer us a rough estimation on the real objective function $\MJ(w)$.
Specifically, if we already known the linear approximation $\tilde{\MJ}(w)$ for a given parameter vector $w$, then the real objective function $\MJ(w)$ is most likely to lie in $\tilde{\MJ}(w) \pm 4\sigma$ (see figure \ref{fig_real_vs_approx_iter0}).
If we further extend the range to $\tilde{\MJ}(w) \pm 6\sigma$ or even to $\tilde{\MJ}(w) \pm 10\sigma$, then the real objective function $\MJ(w)$ should lie in the predicted range.
In another word, if our linear regression is somehow accurate, then it is very unlikely that $\MJ(w)$ falls beyond $\tilde{\MJ}(w) \pm 10\sigma$, i.e.
\begin{equation}
\begin{split}
\MP (\MJ(w) > \MC | \tilde{\MJ}(w) < \MC - 10\sigma) &\rightarrow 0 \\
\MP (\MJ(w) > \MC | \tilde{\MJ}(w) > \MC + 10\sigma) &\rightarrow 1 \\
\end{split}
\end{equation}

Hence, rather than evenly distributing 1,000 strata from 0 to 1, we need to focus on $\tilde{\MJ}(w) \in (\MC - 10\sigma, \MC + 10\sigma)$.
In order to quantitatively estimate the conditional probability $P_i^{(2)}$ we assume the error between $\MJ(w)$ and $\tilde{\MJ}(w)$ has a Laplace distribution.
\begin{equation}
\MJ(w)   = \tilde{\MJ}(w) + \epsilon
\hspace{0.5cm} \text{ where } \hspace{0.5cm}
\epsilon \sim \text{Laplace}(\mu,b)
\end{equation}

the probability density function is given by,
\begin{equation}
f(x|\mu,b) = \frac{1}{2b} \exp \bigg( - \frac{|x-\mu|}{b} \bigg)
\end{equation}
the mean $\mu = 0$ and the variance $2b^2$ should be identical to $\sigma^2$, i.e. $b = \sigma / \sqrt{2}$.
Consider a infinitesimal small stratum $S$ defined as,
\begin{equation}
S = \{w| a - \delta < \tilde{\MJ}(w) < a + \delta \}
\end{equation}

where $a$ can be interpreted the "midpoint" of stratum $S$.
The conditional probability can be estimated as,
\begin{equation}
\MP (\MJ(w) > \MC | w \in S)
\approx \int_{\MC-a}^{+\infty} f(x|\mu,b) dx
= F(+\infty) - F(\MC-a)
\end{equation}

where the cumulative distribution function $F(x)$ is given by,
\begin{equation}
F(x) = \int_{-\infty}^{x} f(u) du = \frac{1}{2} + \frac{1}{2} \sgn(x-\mu) \bigg[1-\exp\bigg(-\frac{|x-\mu|}{b}\bigg)\bigg]
\end{equation}

Thus,
\begin{equation}
\begin{split}
\MP (\MJ(w) > \MC | w \in S)
&\approx 1 - \bigg\{ \frac{1}{2} + \frac{1}{2} \sgn(\MC-a-\mu)\bigg[1-\exp\bigg(-\frac{|\MC-a-\mu|}{b}\bigg)\bigg]\bigg\} \\
&= \frac{1}{2} - \frac{1}{2} \sgn(\MC-a)\bigg[1-\exp\bigg(-\frac{|\MC-a|}{\sigma/\sqrt{2}}\bigg)\bigg]
\end{split}
\end{equation}
where we used $\mu = 0$ and $b=\sigma/\sqrt{2}$.

\bigskip

If $a > \MC$ i.e. $\MC - a < 0$, we have $S$ to the right of $\MC$
\begin{equation}\label{eqn_P2_right}
\begin{split}
\MP (\MJ(w) > \MC | w \in S)
&= \frac{1}{2} + \frac{1}{2}\bigg[1-\exp\bigg(\frac{\MC-a}{\sigma/\sqrt{2}}\bigg)\bigg] \\
&= 1 - \frac{1}{2} \exp\bigg(\frac{\MC-a}{\sigma/\sqrt{2}}\bigg)
\end{split}
\end{equation}

If $a < \MC$ i.e. $\MC - a > 0$, we have $S$ to the left of $\MC$ (majority of the sampling points)
\begin{equation}\label{eqn_P2_left}
\begin{split}
\MP (\MJ(w) > \MC | w \in S)
&= \frac{1}{2} - \frac{1}{2}\bigg[1-\exp\bigg(\frac{a-\MC}{\sigma/\sqrt{2}}\bigg)\bigg] \\
&= \frac{1}{2} \exp\bigg(\frac{a-\MC}{\sigma/\sqrt{2}}\bigg)
\end{split}
\end{equation}

Following this idea, we generated 100 strata within $\tilde{\MJ}(w)\in (\MC-10\sigma, \MC+10\sigma)$.
Combine with another 2 strata $\tilde{\MJ}(w) \in (-\infty, \MC-10\sigma)$ and $\tilde{\MJ}(w) \in (\MC+10\sigma,+\infty)$, we have $N_S = 102$ strata in total.

\begin{figure}[h!]
  \centering
  \includegraphics[height=7cm,trim=0cm 0.5cm 0cm 1.5cm,clip]{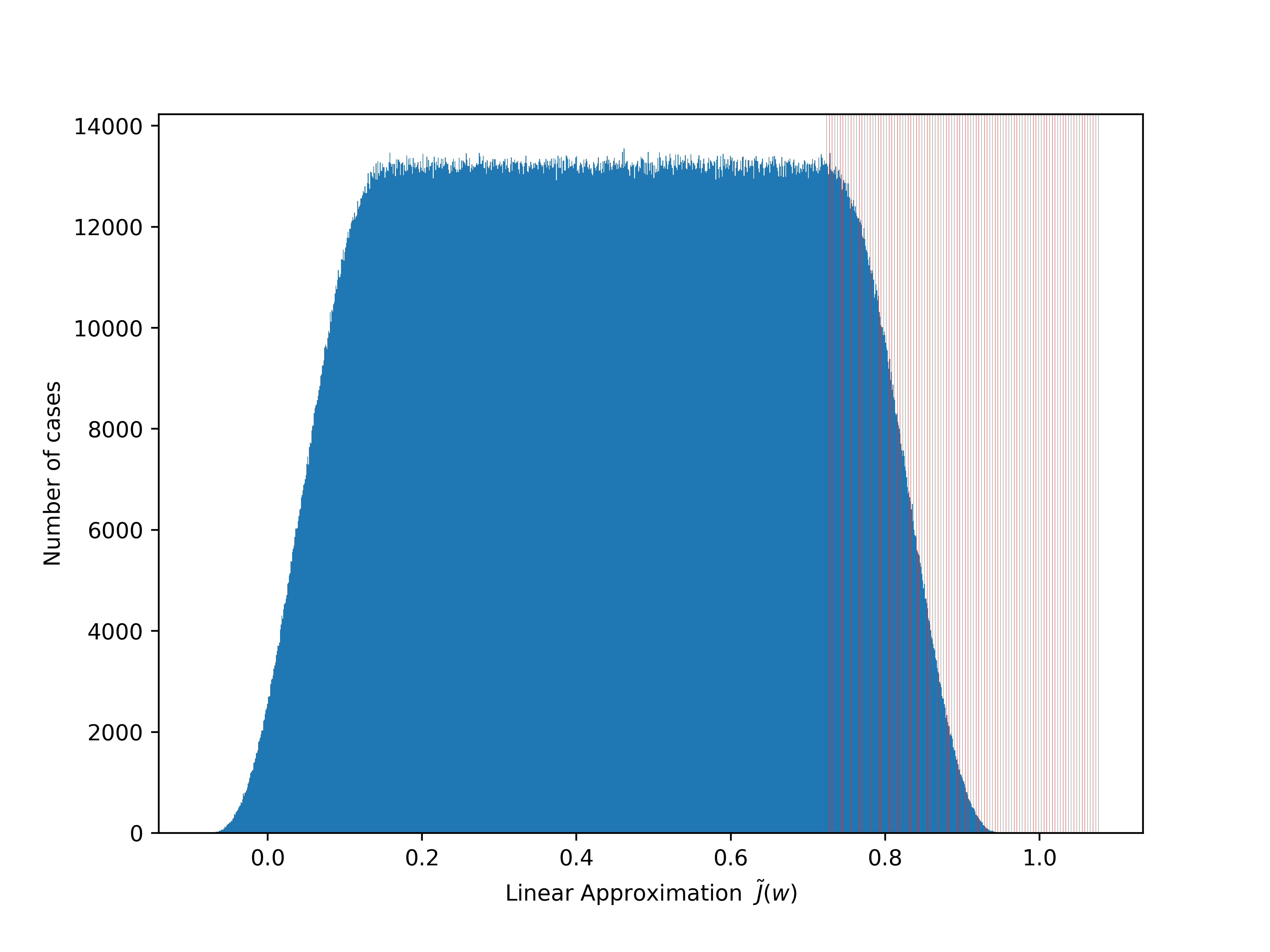}
  \caption{Histogram of linear approximations $\tilde{\MJ}(w)$ with 1000 bins. The vertical red lines indicates the boundary of $N_S = 102$ strata.}
  \label{fig_ApproxJ_hist_iter0}
\end{figure}
\FloatBarrier

\begin{figure}[h!]
  \centering
  \includegraphics[height=6cm,trim=0cm 0.5cm 0cm 1.5cm,clip]{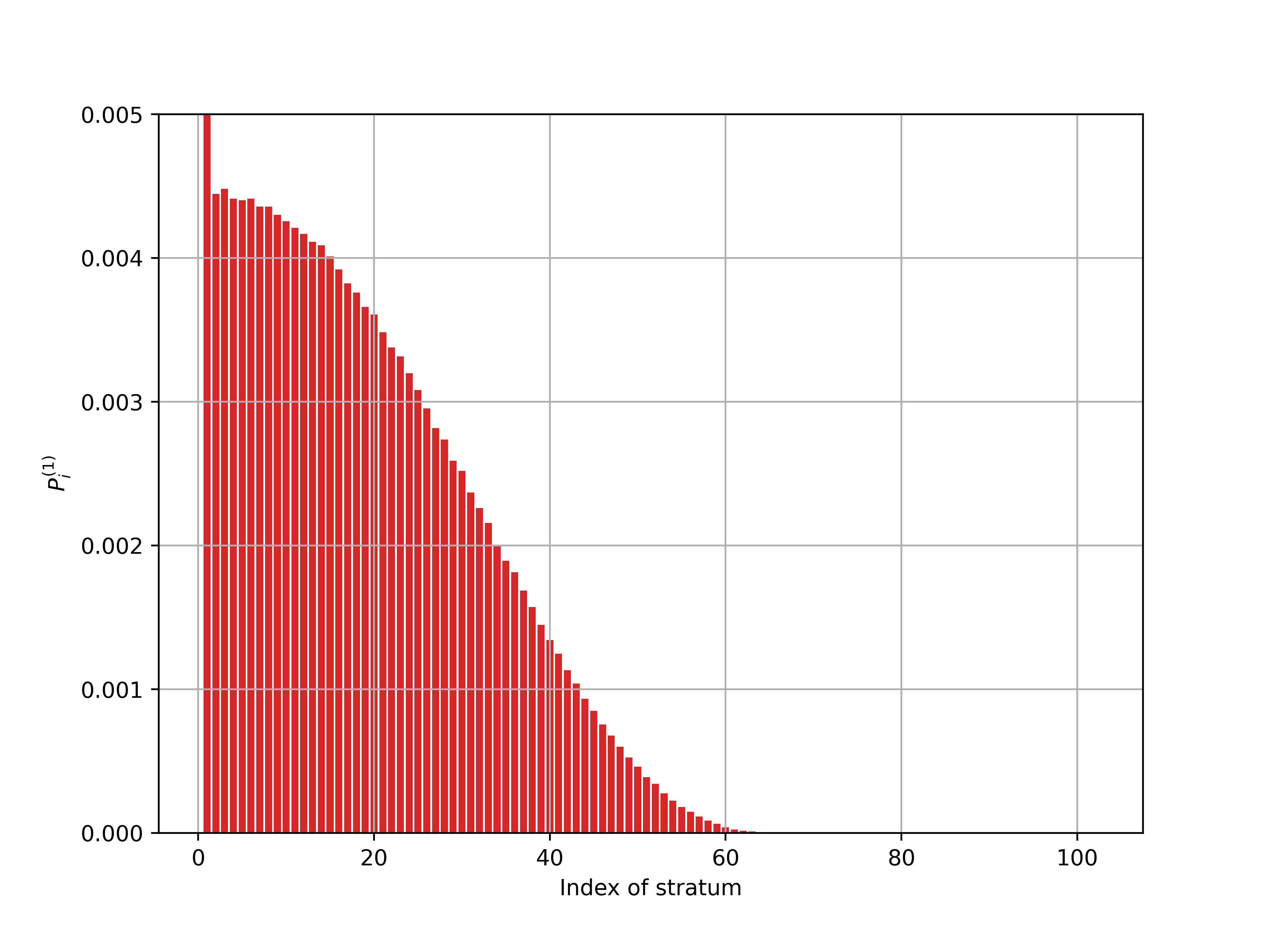}
  \caption{Distribution of $P_i^{(1)}$ with $i = 1 \cdots N_S$. As for the left-most stratum, $P_1^{(1)} \approx 0.8624$.}
  \label{fig_P1_hist_iter0}
\end{figure}

The probability of a sample lies in the $\ith$ stratum $P_i^{(1)}$ is shown as in figure \ref{fig_P1_hist_iter0}.
Since $P_1^{(1)} \gg P_2^{(1)} \cdots P_{N_S}^{(1)}$, we re-scale the y-axis to better illustrate $P_2^{(1)} \cdots P_{N_S}^{(1)}$.
\FloatBarrier

\begin{figure}[h!]
  \centering
  \includegraphics[height=6cm,trim=1cm 0.5cm 2cm 1.5cm,clip]{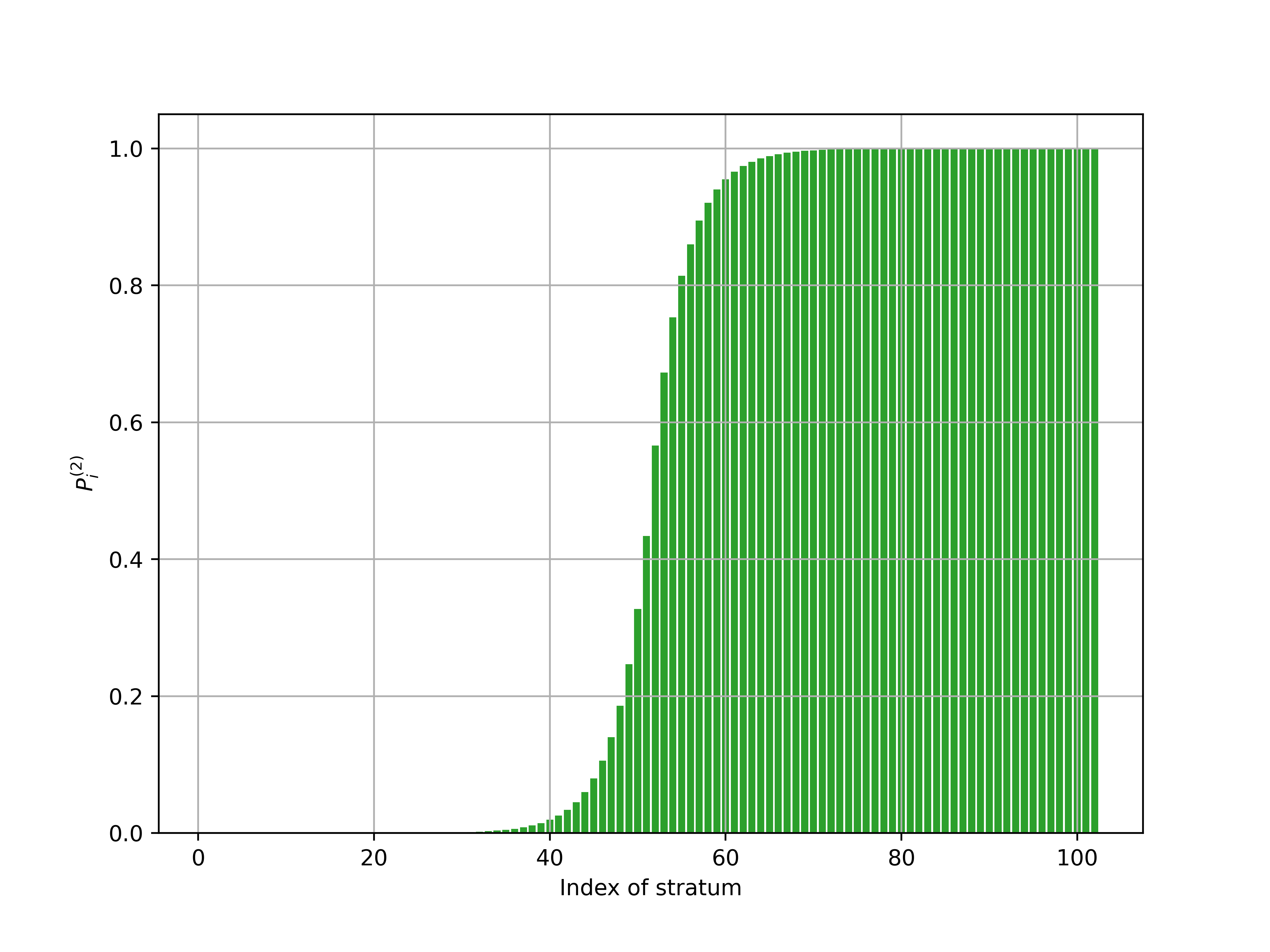}
  \includegraphics[height=6cm,trim=1cm 0.5cm 2cm 1.5cm,clip]{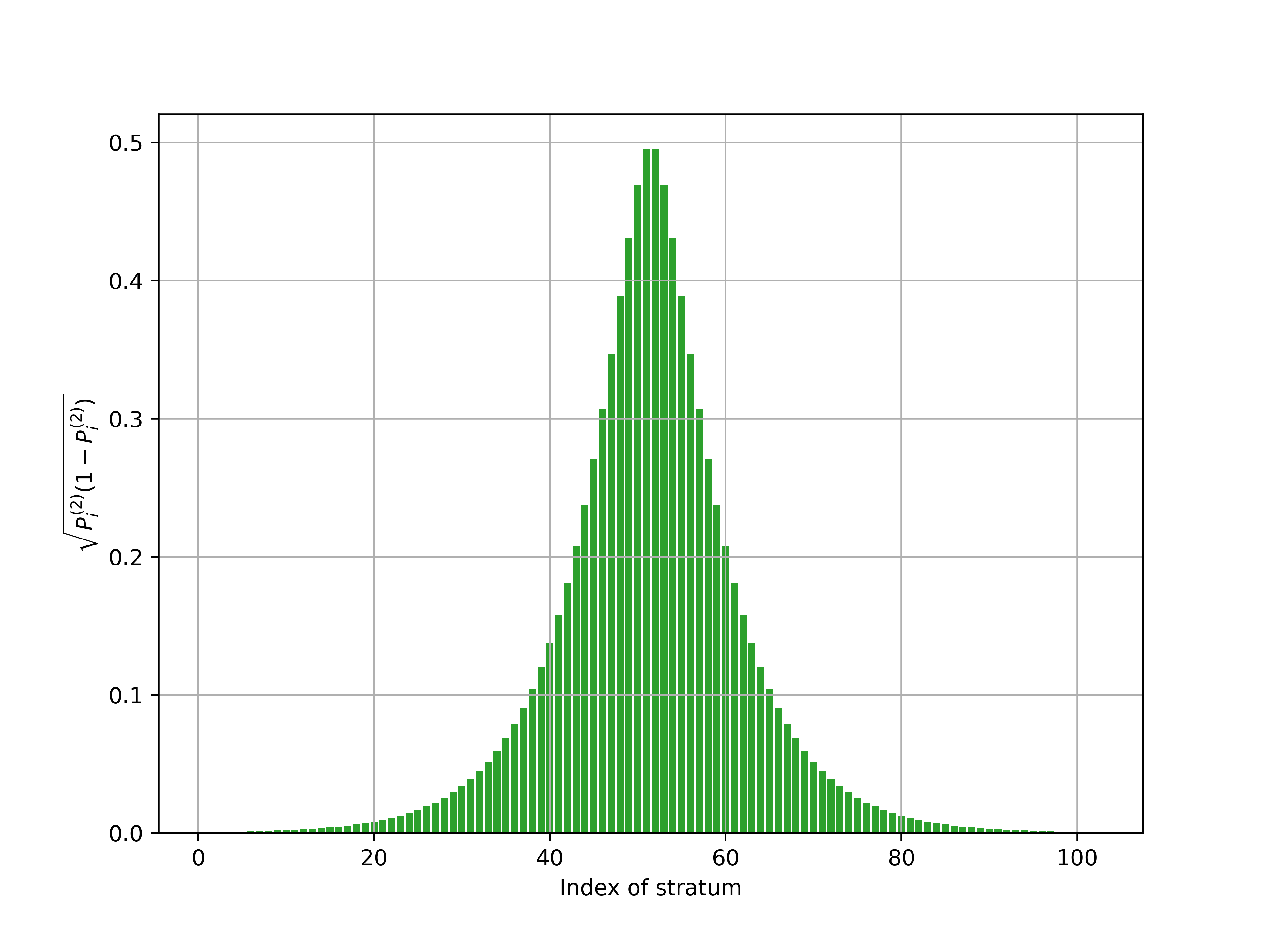}
  \caption{Left: Distribution of $P_i^{(2)}$.
           Right: Distribution of $\sqrt{P_i^{(2)}\big(1-P_i^{(2)}\big)}$.}
  \label{fig_P2_hist_iter0}
\end{figure}

The conditional probability $P_i^{(2)}$ is plotted in figure \ref{fig_P2_hist_iter0}, where we assume that $P_1^{(2)} = 0$ for the left-most strata and $P_{N_s}^{(2)} = 1$ for the right-most strata.
$P_i^{(2)}$ for $i = 2 \cdots N_s - 1$ are calculated from equation (\ref{eqn_P2_right}) and equation (\ref{eqn_P2_left}).

\FloatBarrier

\begin{figure}[h!]
  \centering
  \includegraphics[height=6cm,trim=0cm 0.5cm 2cm 1.5cm,clip]{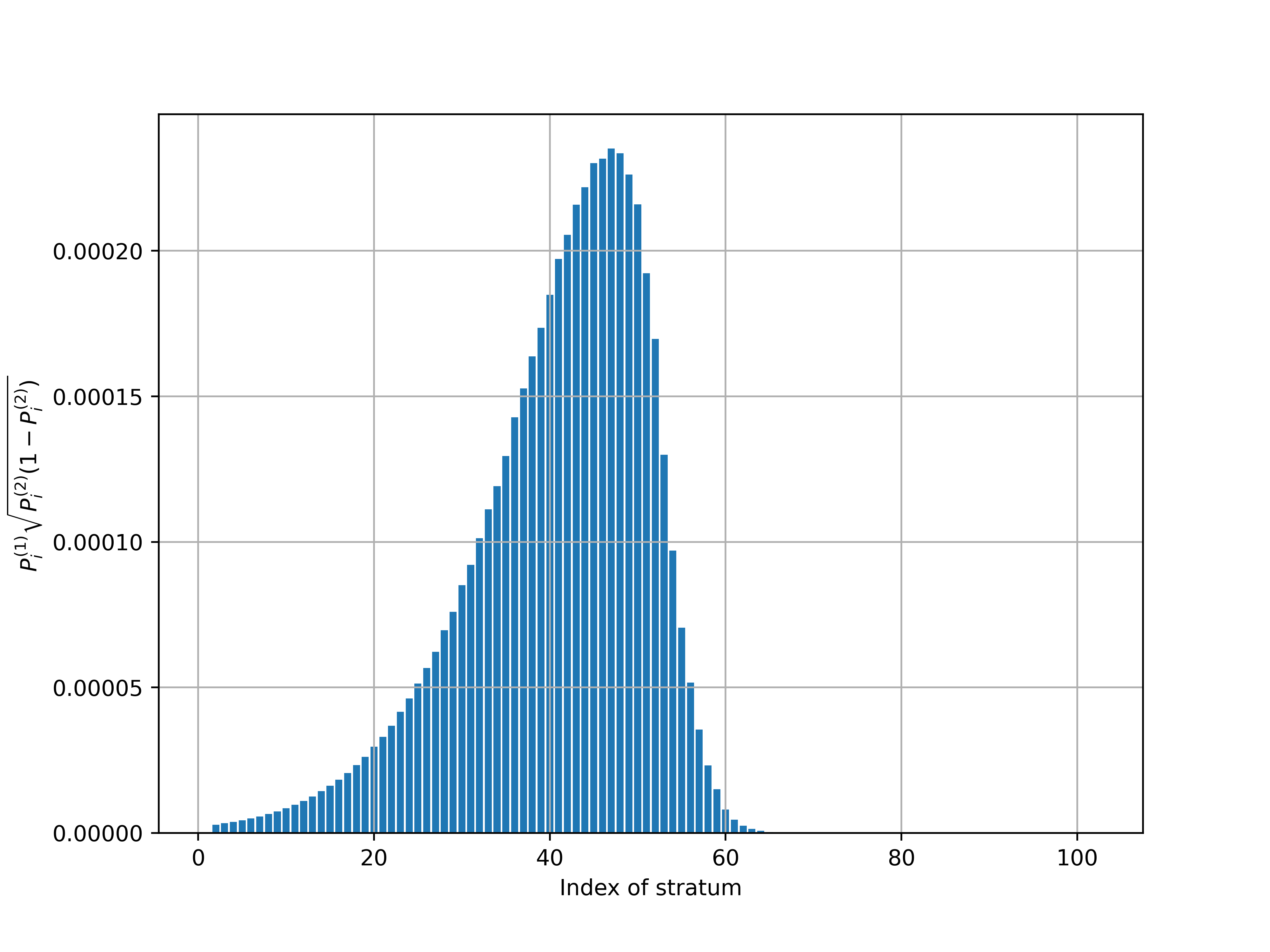}
  \includegraphics[height=6cm,trim=1cm 0.5cm 2cm 1.5cm,clip]{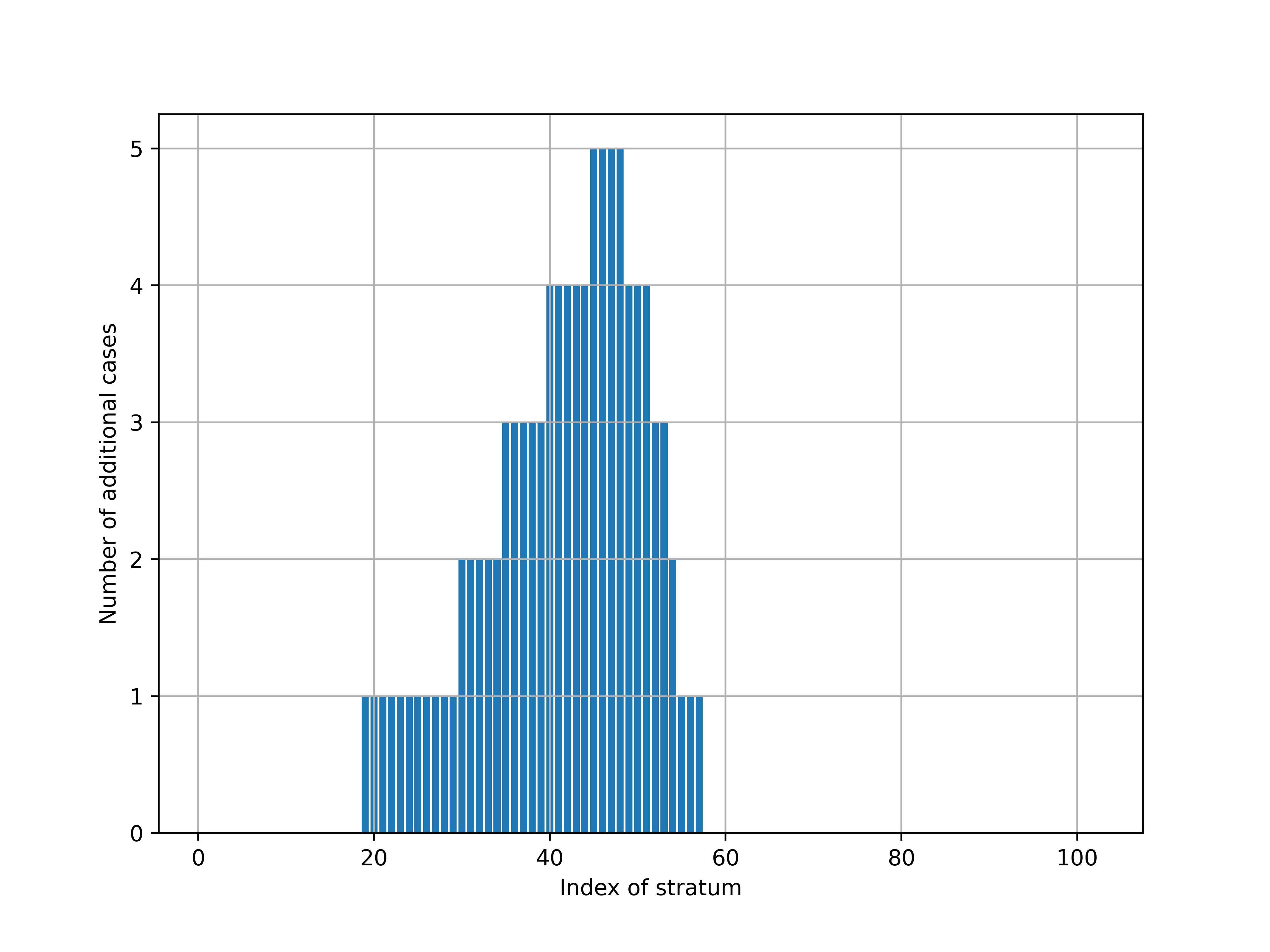}
  \caption{Left: Distribution of $P_i^{(1)} \sqrt{P_i^{(2)}\big(1 - P_i^{(2)}\big)}$.
           Right: Distribution of $N_i$ for the $\ith$ stratum $S_i$, where $\sum N_i = 99$}
  \label{fig_Ni_hist_iter0}
\end{figure}
\FloatBarrier

Combine figure \ref{fig_P1_hist_iter0} and figure \ref{fig_P2_hist_iter0}, the distribution of $P_i^{(1)} \sqrt{P_i^{(2)}\big(1 - P_i^{(2)}\big)}$ as well as the number of additional sampling points are depicted in figure \ref{fig_Ni_hist_iter0}.

\section{Additional Cases with Optimized Allocation}
\begin{figure}[h!]
  \centering
  \includegraphics[height=7cm,trim=0cm 0.5cm 0cm 1.5cm,clip]{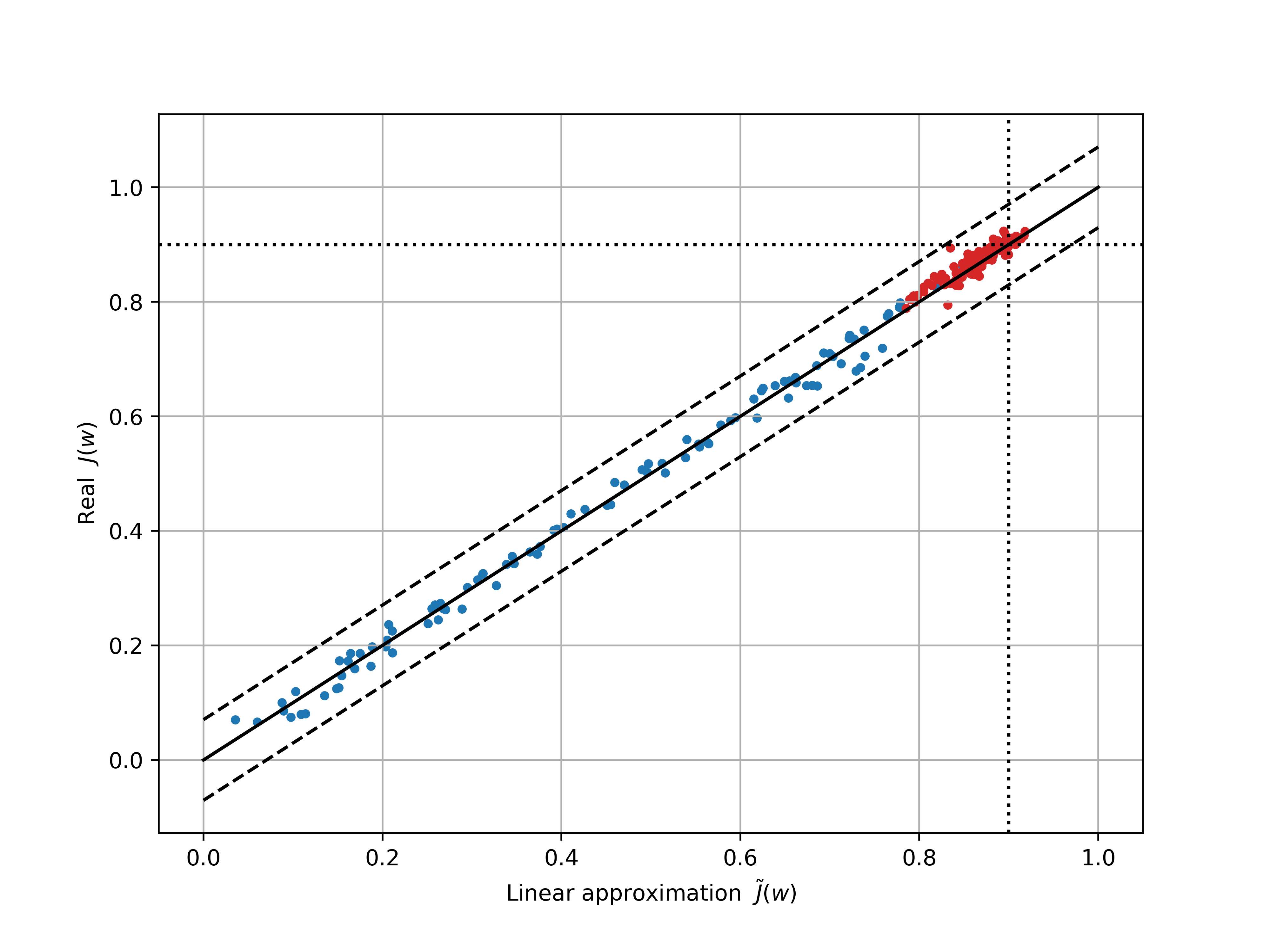}
  \caption{Real objective function $\MJ(w)$ versus the linear approximation $\tilde{\MJ}(w)$. The blue dots indicate the 100 preliminary solutions whereas the red dots show the additional 99 cases.}
  \label{fig_real_vs_approx_iter1}
\end{figure}

Following the $N_i$ distribution plotted in figure \ref{fig_Ni_hist_iter0}, we run additional 99 flow simulations, distributed from stratum 19 to stratum 57.
In each stratum $S_i$, the first $N_i$ samples are selected to determine the geometric and freestream parameters.
Similar to figure \ref{fig_real_vs_approx_iter0}, we plot the real objective function $\MJ(w)$ against the linear approximation $\tilde{\MJ}(w)$ in figure \ref{fig_real_vs_approx_iter1},
\FloatBarrier

It is worth noting that the linear regression was created based on the 100 preliminary flow solutions.
In other words, the 100 preliminary cases and the 99 additional cases can be interpreted as the training data set and the testing data set, respectively.
As shown in figure \ref{fig_real_vs_approx_iter1}, the additional 99 cases are in accordance with the prediction and further validated our linear regression.

\bigskip

\begin{figure}[h!]
  \centering
  \includegraphics[height=6cm,trim=0.75cm 0.5cm 2cm 1.5cm,clip]{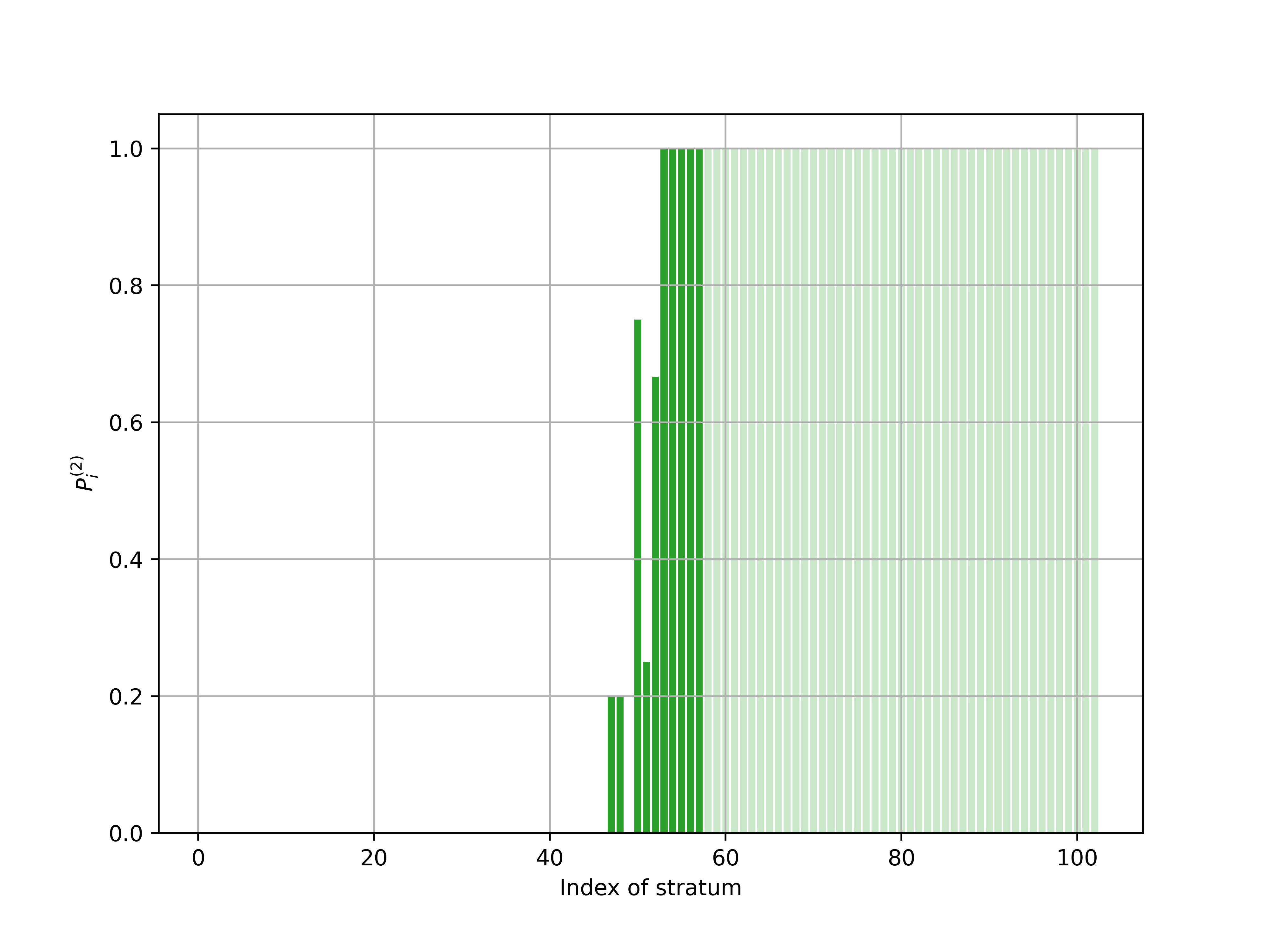}
  \includegraphics[height=6cm,trim=0.75cm 0.5cm 2cm 1.5cm,clip]{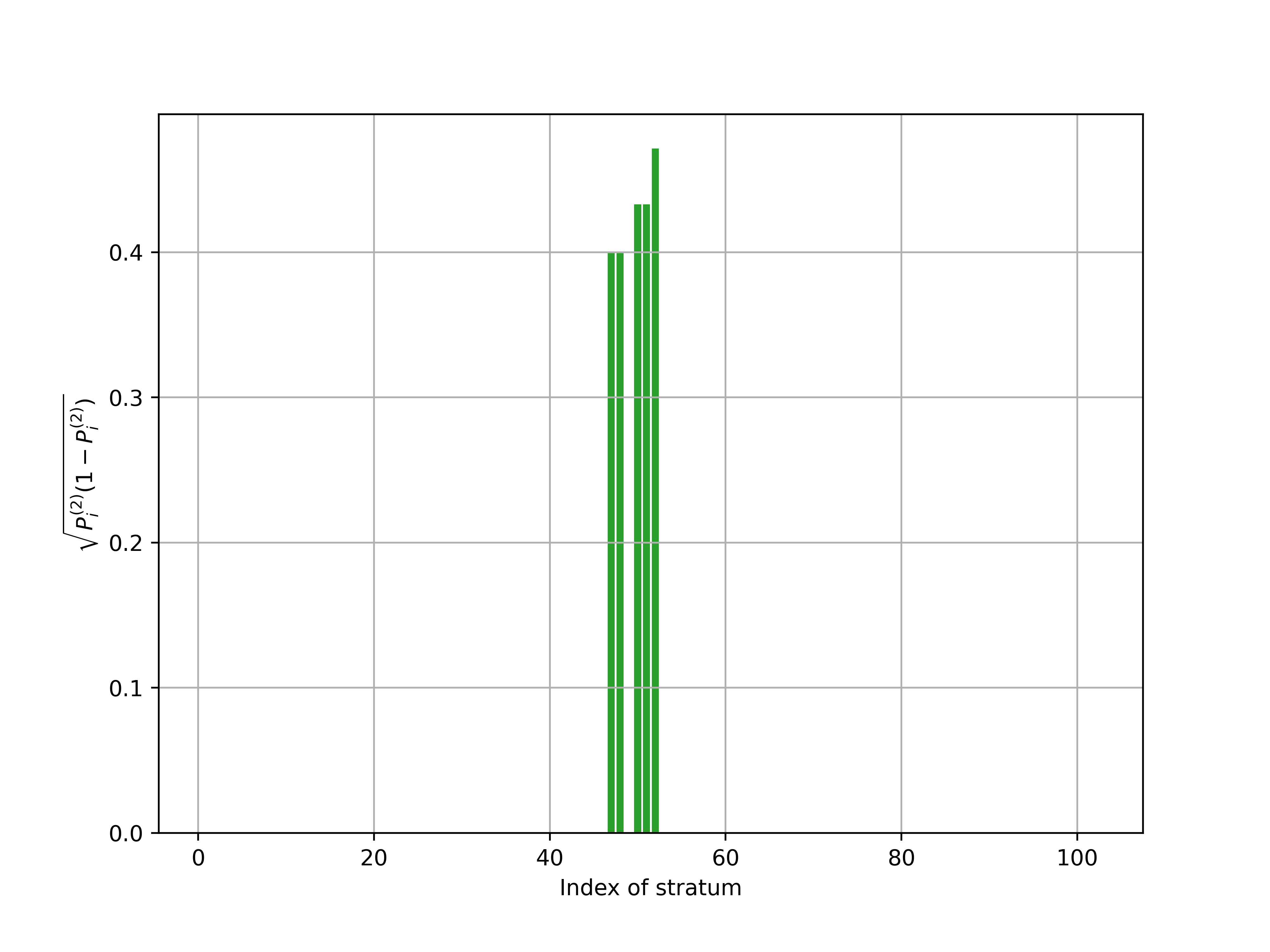}
  \caption{Left: Distribution of $P_i^{(2)}$, the solid bars are real flow solutions while the shaded bars are extrapolation.
  Right: Distribution $\sqrt{P_i^{(2)}\big(1-P_i^{(2)}\big)}$.}
  \label{fig_P2_hist_iter1}
\end{figure}

Now we have 199 flow solutions in total, randomly distributed from stratum 1 to stratum 57.
Majority of the 100 preliminary flow solutions are in the first stratum $S_1 = \{w|\tilde{\MJ}(w) < 0.9 - 10 \sigma\}$,
while the additional 99 cases are spread from stratum 19 to stratum 57.
We want to estimate,
\begin{equation}
\MP(\MJ(w) > \MC) = \sum_{i=1}^{N_S} \MP (w \in S_i) \MP (\MJ(w) > \MC | w \in S_i)
\end{equation}
where $\MP (w \in S_i) \approx P_i^{(1)}$ has already been accurately estimated based on 10,000,000 linear approximations.
Hence we only need to estimate $P_i^{(2)}$ utilizing our 199 flow solutions.
The conditional probability $P_i^{(2)}$ and $\sqrt{P_i^{(2)}\big(1-P_i^{(2)}\big)}$ are depicted in figure \ref{fig_P2_hist_iter1}.
Note that we assume $P_i^{(2)} = 0$ for strata 1 - 18 and $P_i^{(2)} = 1$ for strata 58 - 102 as no sampling point lies in these strata.
Combine $P_2^{(1)}$ depicted in figure \ref{fig_P2_hist_iter1} with $P_i^{(1)}$ plotted in figure \ref{fig_P1_hist_iter0}, the probability $\MP(\MJ(w)>\MC)$ can be estimated as,
\begin{equation}
\begin{split}
\MP(\MJ(w) > \MC)
&= \sum_{i=1}^{N_S} \MP (w \in S_i) \MP (\MJ(w) > \MC | w \in S_i) \\
&\approx \sum_{i=1}^{N_S} P_i^{(1)} P_i^{(2)} \approx 0.00213
\end{split}
\end{equation}

\FloatBarrier

The biased sample variance is given by,
\begin{equation}
\begin{split}
\Var \bigg[\sum_{i=1}^{N_S} P_i^{(1)} P_i^{(2)}\bigg]
&= \sum_{i=1}^{N_S} \big(P_i^{(1)}\big)^2 \Var \big[ P_i^{(2)} \big]
= \sum_{i=1}^{N_S} \big(P_i^{(1)}\big)^2 \frac{P_i^{(2)}\big(1-P_i^{(2)}\big)}{N_i} \\
&\approx \text{5.191024e-08} \\
\end{split}
\end{equation}

where we assume $P_i^{(1)}$ are constants rather than random variables.
Furthermore, the unbiased population variance $s^2$ can be written as,
\begin{equation}
\begin{split}
s^2 
&= \sum_{i=1}^{N_S} \big(P_i^{(1)}\big)^2 \big( s_i^{(2)} \big)^2
= \sum_{i=1}^{N_S} \big(P_i^{(1)}\big)^2 \frac{P_i^{(2)}\big(1-P_i^{(2)}\big)}{N_i-1} \\
& \approx \text{6.847554e-08} \\
\end{split}
\end{equation}

where all strata with $N_i = 1$ are neglected.
Since we have $\mu = 0.00213, s = 0.000262$, the 95\% confidence interval is,
\begin{equation}
(\mu-2s, \mu+2s) = (0.00160, 0.00265)
\end{equation}

\subsection{Comparison with Naive Monte Carlo method}
In order to achieve similar level of variance, i.e. width of confidence interval, how many samples do we need if we use naive Monte Carlo method?
Denote the total number of sampling points as $N_A$, the estimated probability of achieving high lift coefficient can be written as,
\begin{equation}
\MP (\MJ(w) > \MC) \approx \frac{1}{N_A} \sum_{k=1}^{N_A} I_{\MJ(w_k) > \MC}
\end{equation}

where,
\begin{equation}
I_{\MJ(w_k) > \MC} =
\begin{cases}
  1, & \mbox{if }\MJ(w_k) > \MC \\
  0, & \mbox{if } \MJ(w_k) < \MC \\
\end{cases}
\end{equation}

The biased sample variance of $\MP (\MJ(w) > \MC)$ is given by,
\begin{equation}
\Var\big[ \MP (\MJ(w) > \MC) \big] 
= \Var \bigg[ \frac{1}{N_A} \sum_{k=1}^{N_A} I_{\MJ(w_k) > \MC} \bigg]
= \frac{1}{N_A} \ \Var\big[ I_{\MJ(w_k) > \MC} \big]
\end{equation}

where,
\begin{equation}
\Var\big[ I_{\MJ(w_k) > \MC} \big] = \MP (\MJ(w) > \MC) \big( 1-\MP (\MJ(w) > \MC) \big)
\end{equation}

Thus,
\begin{equation}
\Var \big[ \MP (\MJ(w) > \MC) \big] = \frac{1}{N_A} \MP (\MJ(w) > \MC) \big(1-\MP (\MJ(w) > \MC)\big)
\end{equation}

which gives,
\begin{equation}
N_A = \frac{\MP (\MJ(w) > \MC) \big(1-\MP (\MJ(w) > \MC)\big)}
           {\Var \big[ \MP (\MJ(w) > \MC) \big]}
\end{equation}

Since we want $\Var \big[ \MP (\MJ(w) > \MC) \big] \approx \text{5.191024e-08}$,
\begin{equation}
N_A \approx 40,852
\end{equation}

In order to achieve the same level of variance obtained by the adaptive sampling approach, we will have to run 40,852 simulations in total when using the naive Monte Carlo method.
However, in the adaptive approach, we only simulated 199 cases in total: 100 preliminary cases plus 99 additional cases.
Therefore, the adaptive sampling approach we proposed is over 200 time faster compared to the naive Monte Carlo method, while providing the exactly same level of accuracy.

\section{Multiple Adaptive Iterations with Smaller Ensemble Size}
As shown in the previous section, the adaptive sampling process is much more efficient compared to the naive Monte Carlo method.
However, is it possible for us to further improve our adaptive sampling method, by running more adaptive iterations with smaller ensemble size $\sum N_i$ per iteration?
In other words, can we achieve similar level of accuracy with less number of sampling points in total?

\begin{figure}[h!]
  \centering
  \includegraphics[height=7cm,trim=0cm 0.5cm 0cm 1.5cm,clip]{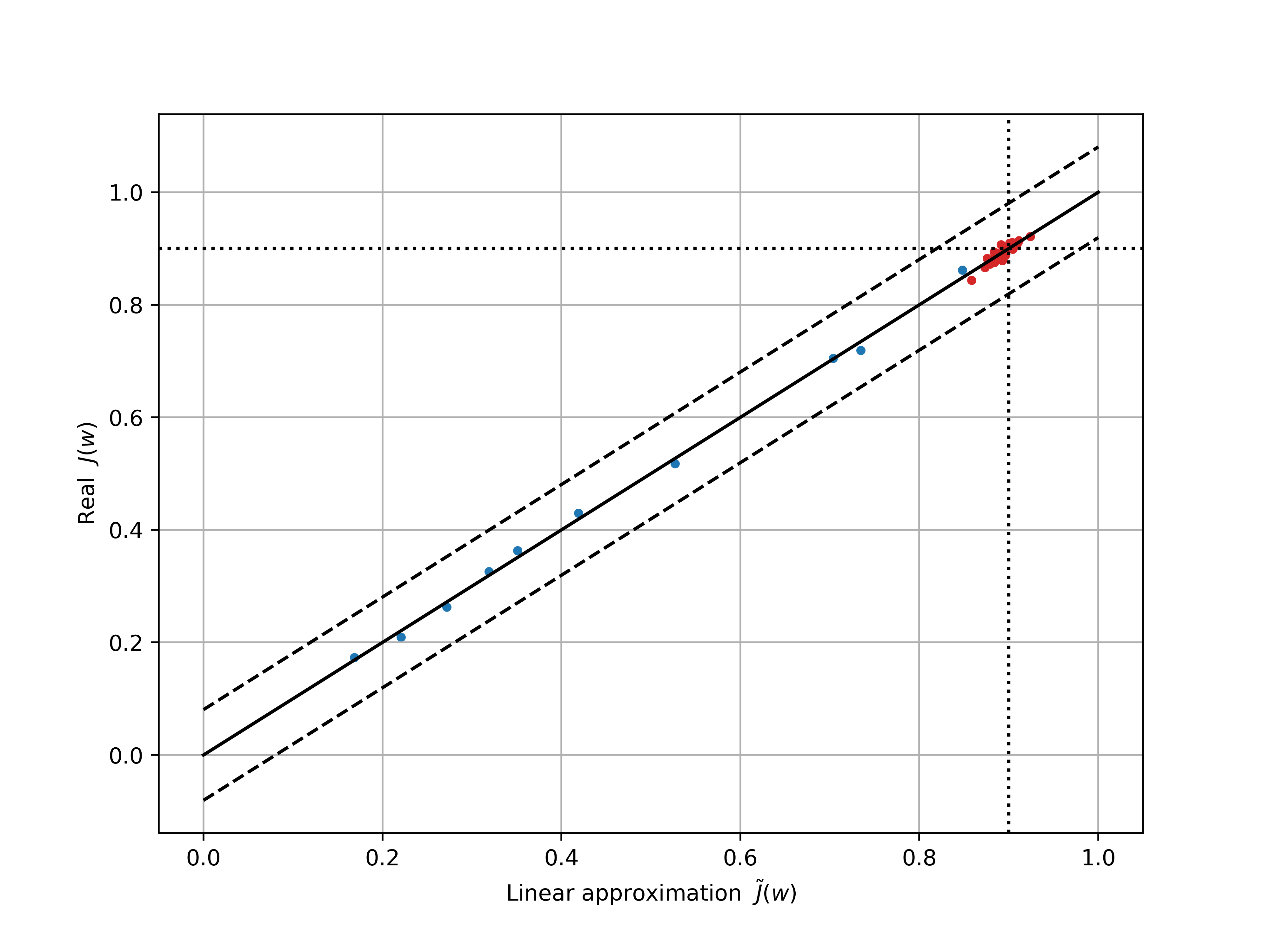}
  \caption{Real objective function $\MJ(w)$ versus the linear approximation $\tilde{\MJ}(w)$. The blue dots indicate the 10 preliminary solutions whereas the red dots show the additional 30 cases.}
  \label{fig_eco_real_vs_approx_iter1_mixed}
\end{figure}
\FloatBarrier

Similar to previous section, we run 10 preliminary cases first, followed by 30 additional cases.
In order to avoid under-sampling, while reducing number of sampling points, we also need to reduce the number of strata.
Hence, we only split $\tilde{\MJ}(w) \in (\MC - 10\sigma, \MC + 10\sigma)$ into 20 strata.
Combined with another 2 strata $\tilde{\MJ}(w) \in (-\infty, \MC - 10\sigma)$ and $\tilde{\MJ}(w) \in (\MC + 10\sigma,+\infty)$, we have $N_S=22$ strata in total.
Instead of spending 100 sampling points to generate an accurate regression model at the very beginning, a better strategy is to keep improving the regression model from iteration to iteration.

\begin{figure}[h!]
  \centering
  \includegraphics[height=6cm,trim=1cm 0.5cm 2cm 1.5cm,clip]{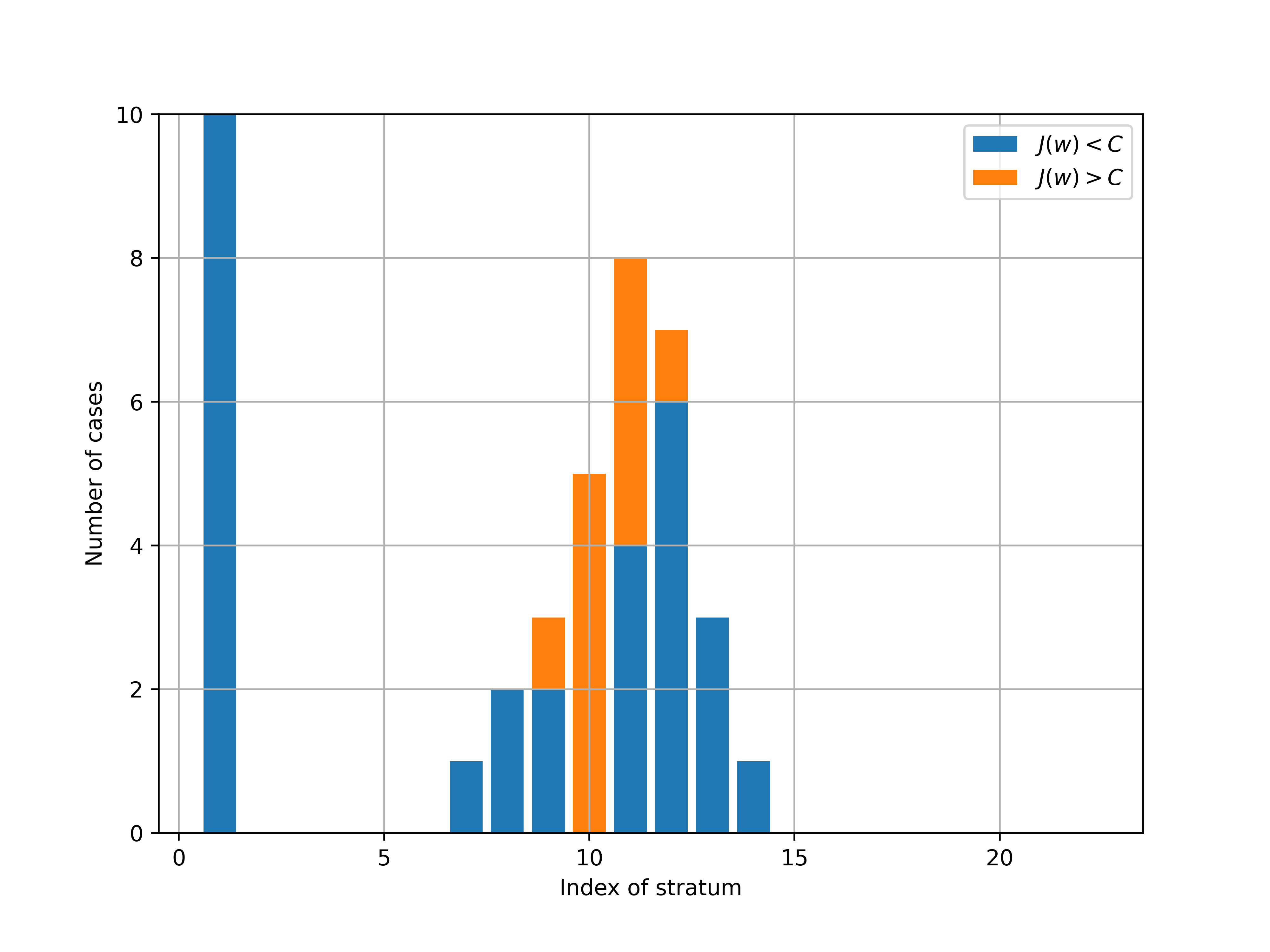}
  \includegraphics[height=6cm,trim=1cm 0.5cm 2cm 1.5cm,clip]{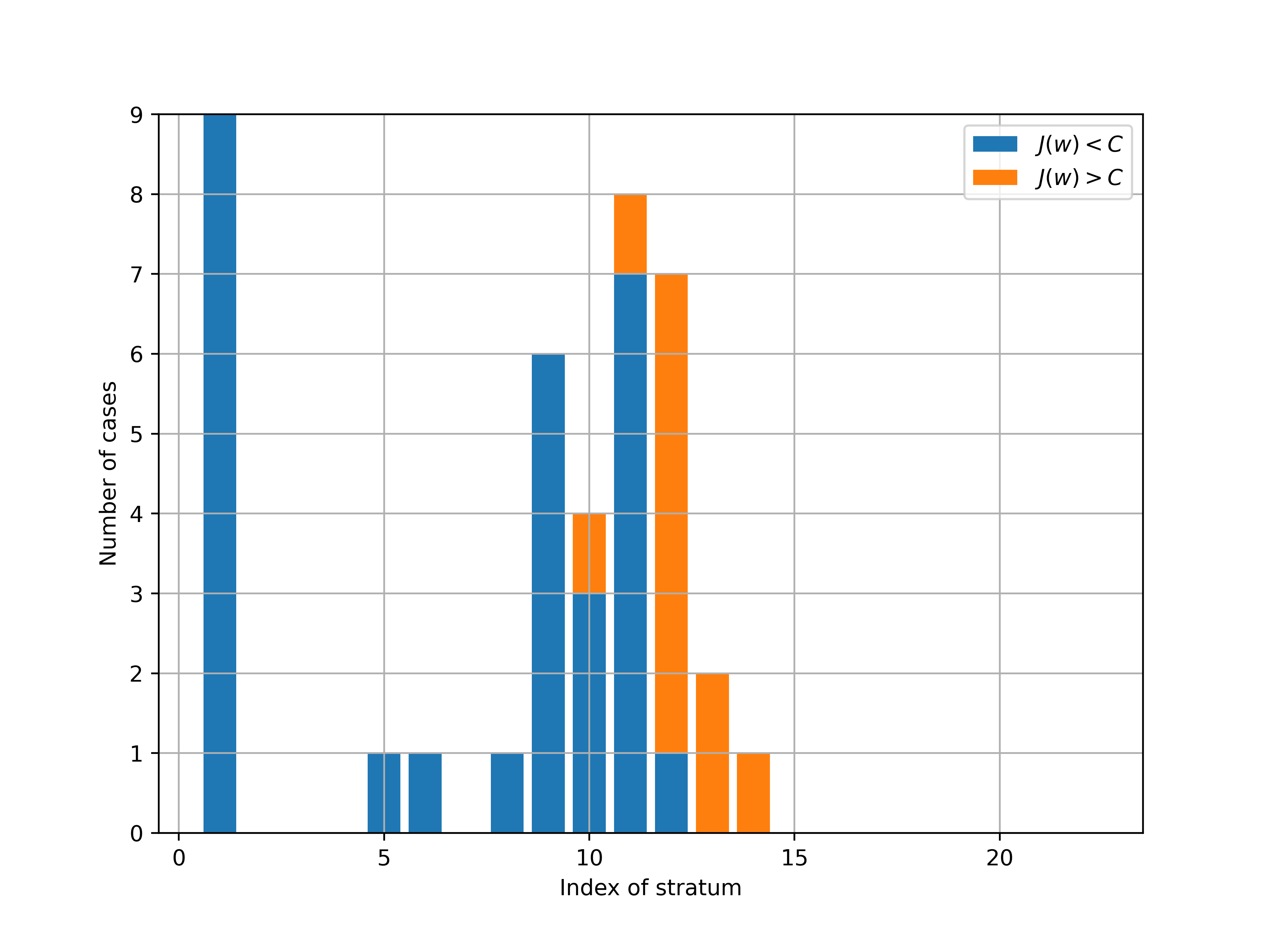}
  \caption{Number of cases distributed in different strata.
  Blue bars: number of cases in $S_i$ with $\MJ(w)<\MC$.
  Orange bars: number of cases in $S_i$ with $\MJ(w)>\MC$.
  Left: regression based on 10 preliminary cases.
  Right: regression based on 10 preliminary + 30 additional cases.}
  \label{fig_eco_NNC_hist_iter1_mixed}
\end{figure}
\FloatBarrier

The left bar plot in figure \ref{fig_eco_NNC_hist_iter1_mixed} shows the number of cases with $\MJ(w)<\MC$ and $\MJ(w)>\MC$ in each stratum,
where the regression model obtained using only 10 preliminary cases is not quite accurate.
Take $S_{10}$ and $S_{13}$ for example:
\begin{itemize}
\item For cases in $S_{10}$, the linear approximated objective functions $\tilde{\MJ}(w)$ are low.
If the regression model were accurate, the real objective $\MJ(w)$ should be low as well.
Hence, the conditional probability $P_{10}^{(2)} = \MP (\MJ(w) > \MC | w \in S_{10})$ should also be low.
However, actually we have $P_{10}^{(2)}=1$.
\item For cases in $S_{13}$, the linear approximations $\tilde{\MJ}(w)$ are high.
Therefore, these cases should be more likely to achieve $\MJ(w) > \MC$,
but we only get $P_{13}^{(2)} \approx 0.143$ which is much lower than $P_{10}^{(2)}=1$.
\end{itemize}

\bigskip
The accuracy of our regression model is greatly improved once we take the 30 additional cases into account.
Using this updated regression model to define the strata, we plot the number of cases in each stratum in the right bar plot.
Higher index of stratum means larger approximated objective function $\tilde{\MJ}(w)$,
which predicts larger real objective function $\MJ(w)$ and hence more likelihood of $\MJ(w) > \MC$.

\begin{figure}[h!]
  \centering
  \includegraphics[height=6cm,trim=0.5cm 0.5cm 2cm 1.5cm,clip]{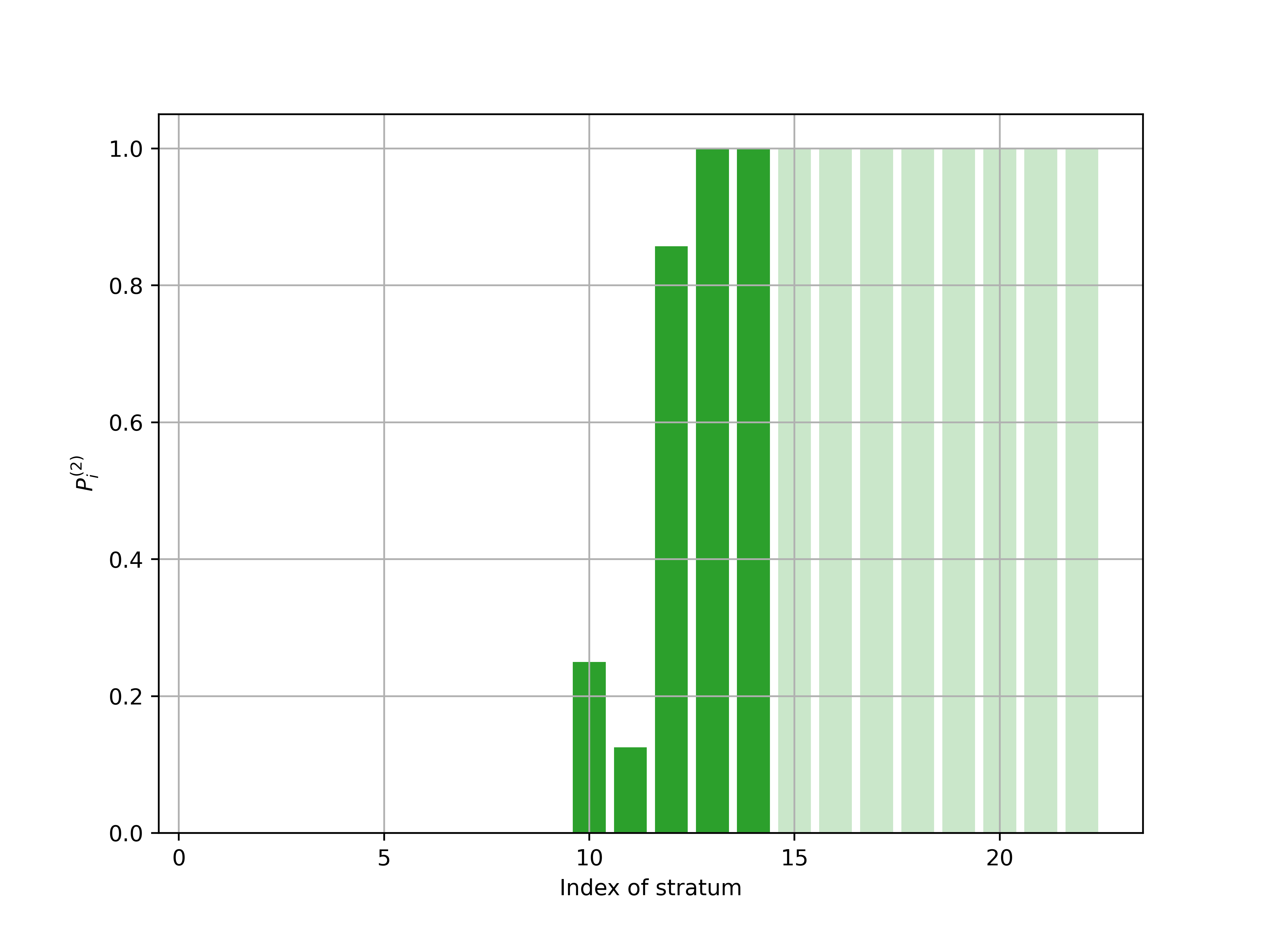}
  \includegraphics[height=6cm,trim=0.5cm 0.5cm 2cm 1.5cm,clip]{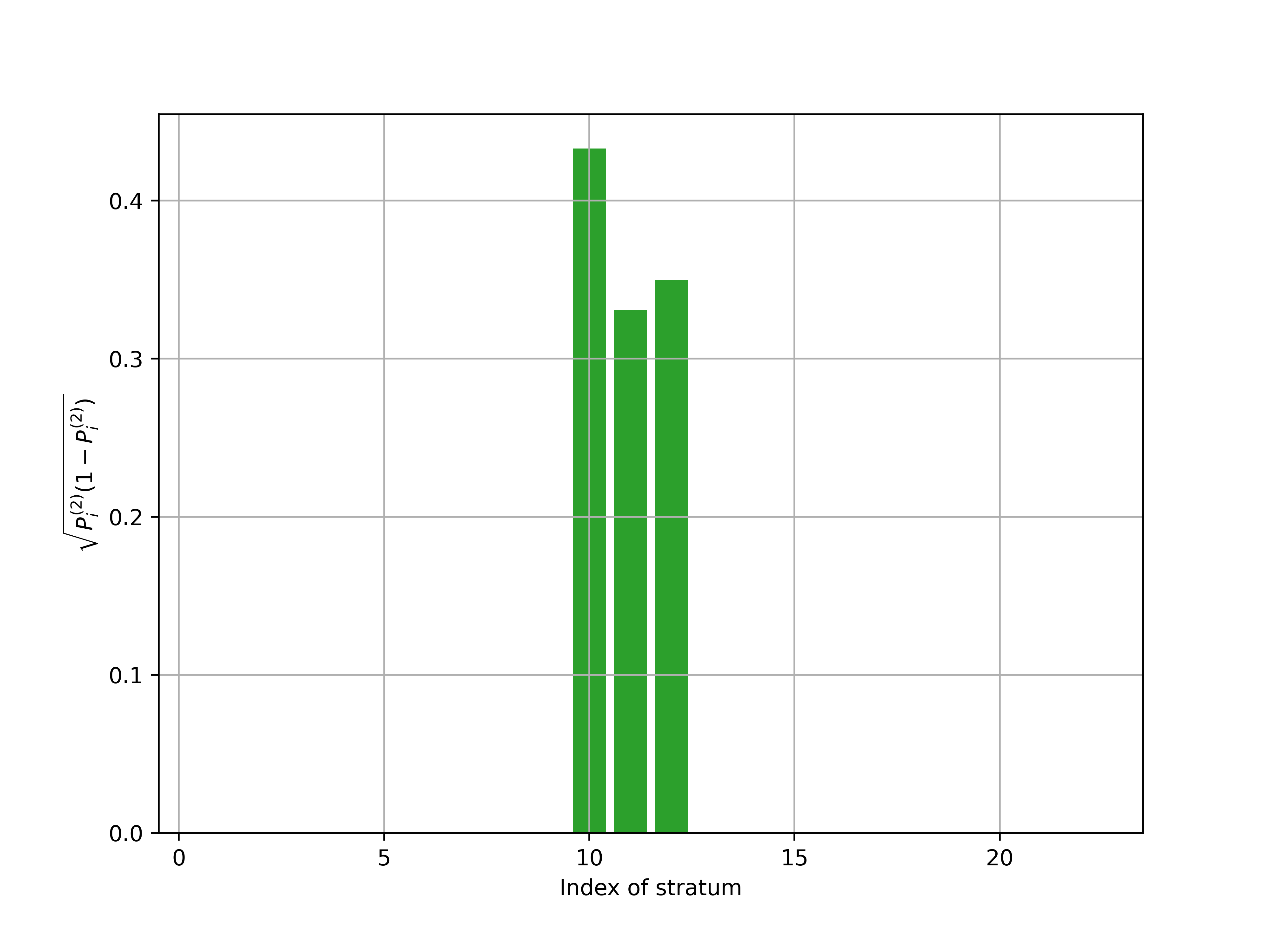}
  \caption{Left: Distribution of $P_i^{(2)}$, the solid bars are real flow solutions while the shaded bars are extrapolation. Right: Distribution of $\sqrt{P_i^{(2)}\big(1-P_i^{(2)}\big)}$.}
  \label{fig_eco_P2_hist_iter1}
\end{figure}
\FloatBarrier

After the first adaptive iteration, $\MP(\MJ(w) > \MC)$ can be estimated as,
\begin{equation}
\MP(\MJ(w) > \MC) \approx 0.00198
\end{equation}

The biased sample variance is given by,
\begin{equation}
\begin{split}
\Var \bigg[\sum_{i=1}^{N_S} P_i^{(1)} P_i^{(2)}\bigg]
&= \sum_{i=1}^{N_S} \big(P_i^{(1)}\big)^2 \frac{P_i^{(2)}\big(1-P_i^{(2)}\big)}{N_i} \\
&\approx \text{8.588410e-08} \\
\end{split}
\end{equation}

The unbiased population variance can be written as,
\begin{equation}
\begin{split}
s^2 &= \sum_{i=1}^{N_S} \big(P_i^{(1)}\big)^2 \frac{P_i^{(2)}\big(1-P_i^{(2)}\big)}{N_i-1} \\
&\approx \text{1.110937e-07} \\
\end{split}
\end{equation}

Since we have $\mu = 0.00198, s = 0.00033$, the 95\% confidence interval is
\begin{equation}
(\mu-2s, \mu+2s) = (0.00131,0.00265) 
\end{equation}

It is worth noting that we only have 10 + 30 cases now, and the variance is just slightly larger than the variance achieved by running 100 + 99 cases previously.
Let us run one more adaptive iteration and surpass the accuracy we previously achieved.

\subsection{Hybrid Model}
If we directly use the distribution of $\sqrt{P_i^{(2)}\big(1-P_i^{(2)}\big)}$ in figure \ref{fig_eco_P2_hist_iter1} to guide the distribution of $N_i$ for the second adaptive iteration, we will have all additional sampling points clustered in $S_{10}$, $S_{11}$ and $S_{12}$, whereas the stratum faraway from $\tilde{\MJ}(w) = 0.9$ will never get sampled.
Therefore, we introduce a hybrid model to mix the probability estimated from the flow solutions (observations) with the probability analytically calculated from the Laplace distribution (predictions).
Specifically,

\bigskip

\begin{itemize}
\item when $N_i=0$, we have to fully depend on the Laplace distribution (prediction).
\item when $N_i$ is large enough, we can 100\% trust the probability estimated from the flow solutions (observations).
\item as for those intermediate $N_i$, we mix the prediction with the observations. The weight is determined by the confidence we have on the observations. Specifically, the more observations we have in stratum $S_i$, the more we trust the $P_i^{(2)}$ estimated from such observations and less rely on the prediction.
\end{itemize}

\bigskip

\begin{figure}[h!]
  \centering
  \includegraphics[height=6cm,trim=0.5cm 0.5cm 2cm 1.5cm,clip]{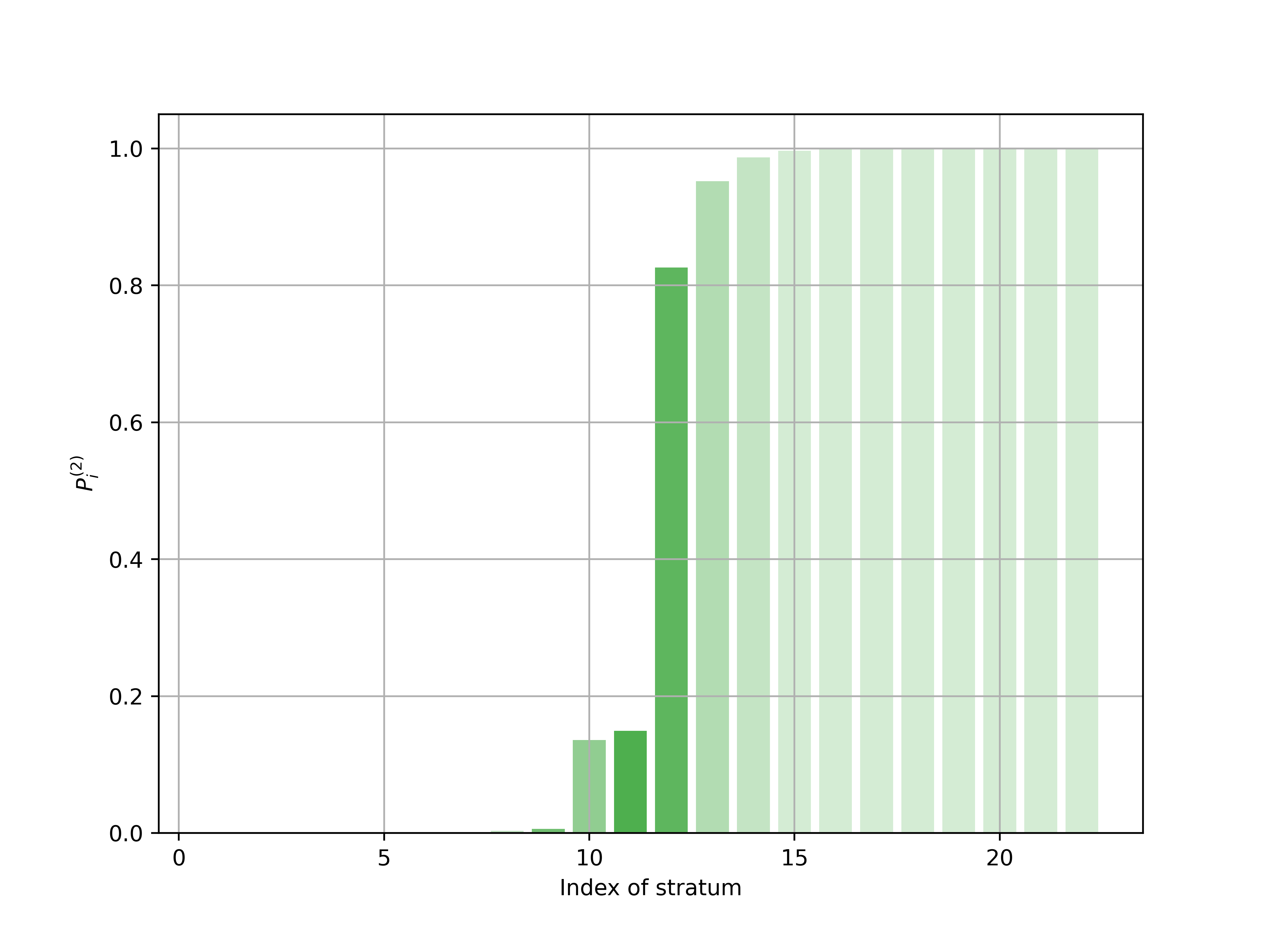}
  \includegraphics[height=6cm,trim=0.5cm 0.5cm 2cm 1.5cm,clip]{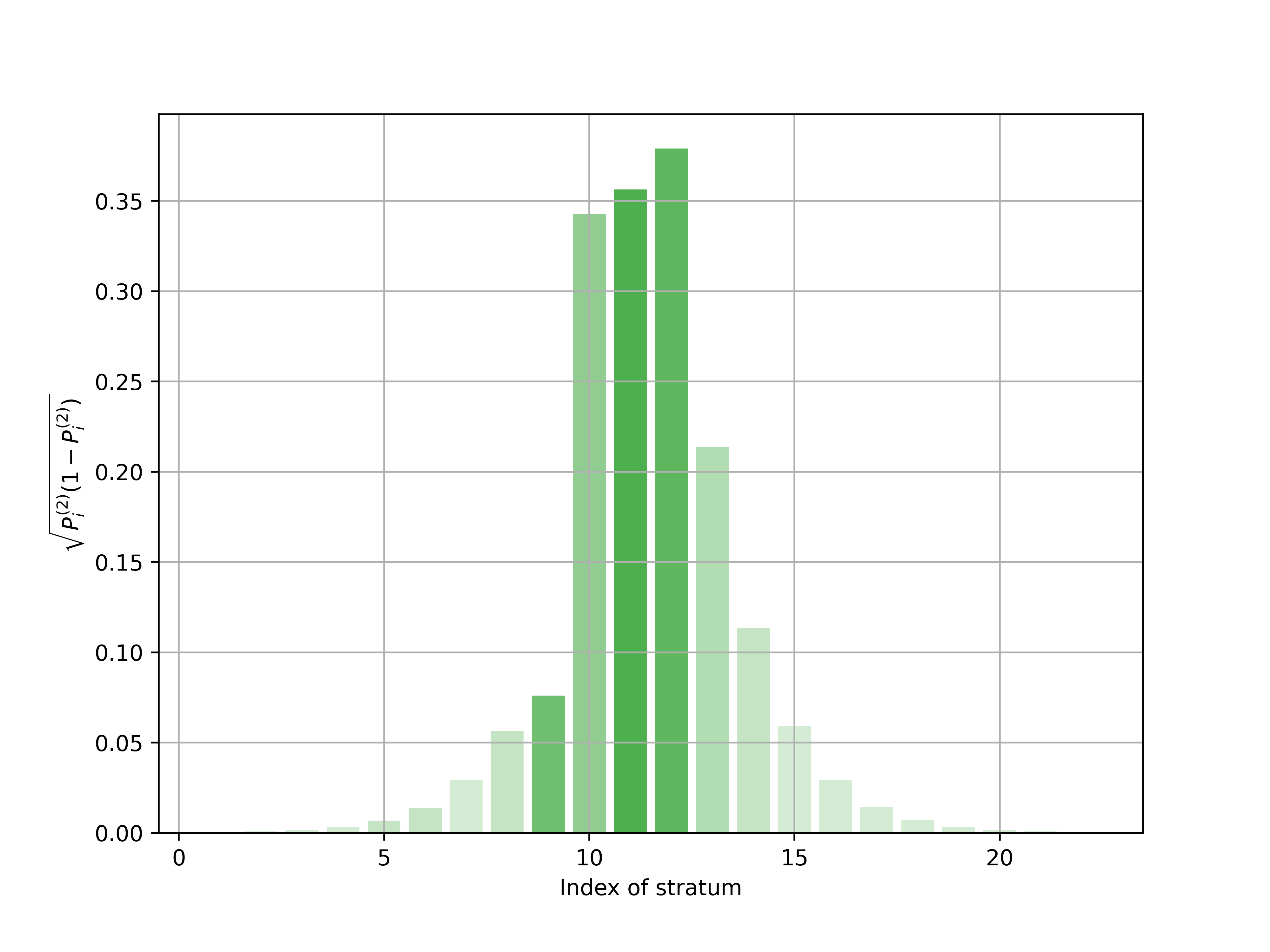}
  \caption{Left: Distribution of $P_i^{(2)}$.
           Right: Distribution of $\sqrt{P_i^{(2)}\big(1-P_i^{(2)}\big)}$.
           The darker bars indicates more flow solutions in $S_i$.}
  \label{fig_eco_P2Mix_hist_iter1}
\end{figure}

Hence we introduce,
\begin{equation}
r_i = \frac{N_i}{N_\text{confident}}
\end{equation}
For instance, we set $N_{\text{confident}} = 10$, i.e. if we have 10 (or more) samples per stratum, then we can 100\% trust the $P_i^{(2)}$ estimated from the flow solutions.
\begin{equation}
P_{i,\text{mix}}^{(2)} = r_i P_{i,\text{observation}}^{(2)} + (1-r_i) P_{i,\text{prediction}}^{(2)}
\end{equation}
By introducing this idea of mixing the prediction with observations, we will be able to sample every stratum after running adequate adaptive sampling iterations.
The mixed $P_i^{(2)}$ and $\sqrt{P_i^{(2)}\big(1-P_i^{(2)}\big)}$ are depicted in figure \ref{fig_eco_P2Mix_hist_iter1},
\FloatBarrier

\begin{figure}[h!]
  \centering
  \includegraphics[height=6cm,trim=0cm 0.5cm 2cm 1.5cm,clip]{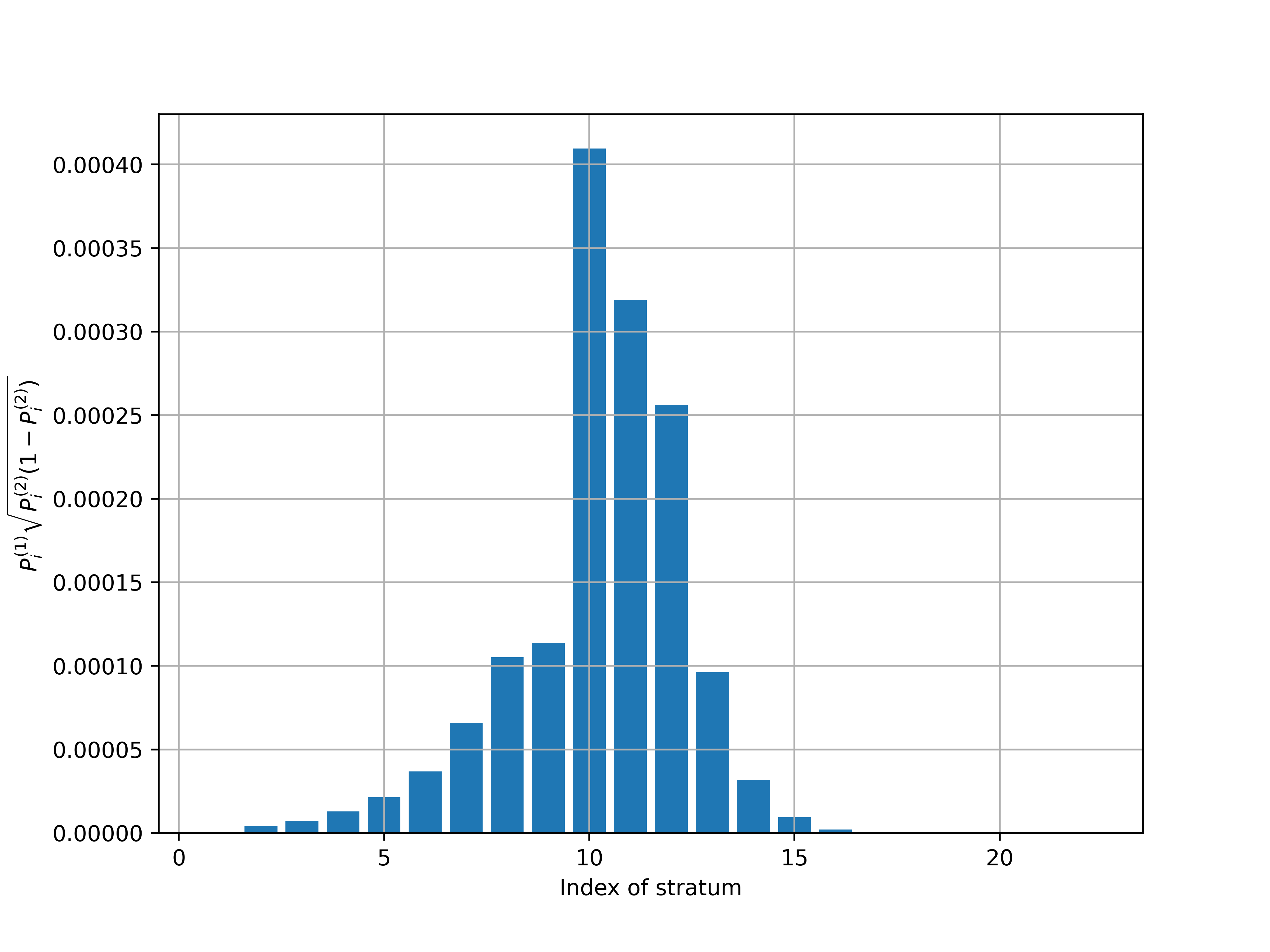}
  \includegraphics[height=6cm,trim=1cm 0.5cm 2cm 1.5cm,clip]{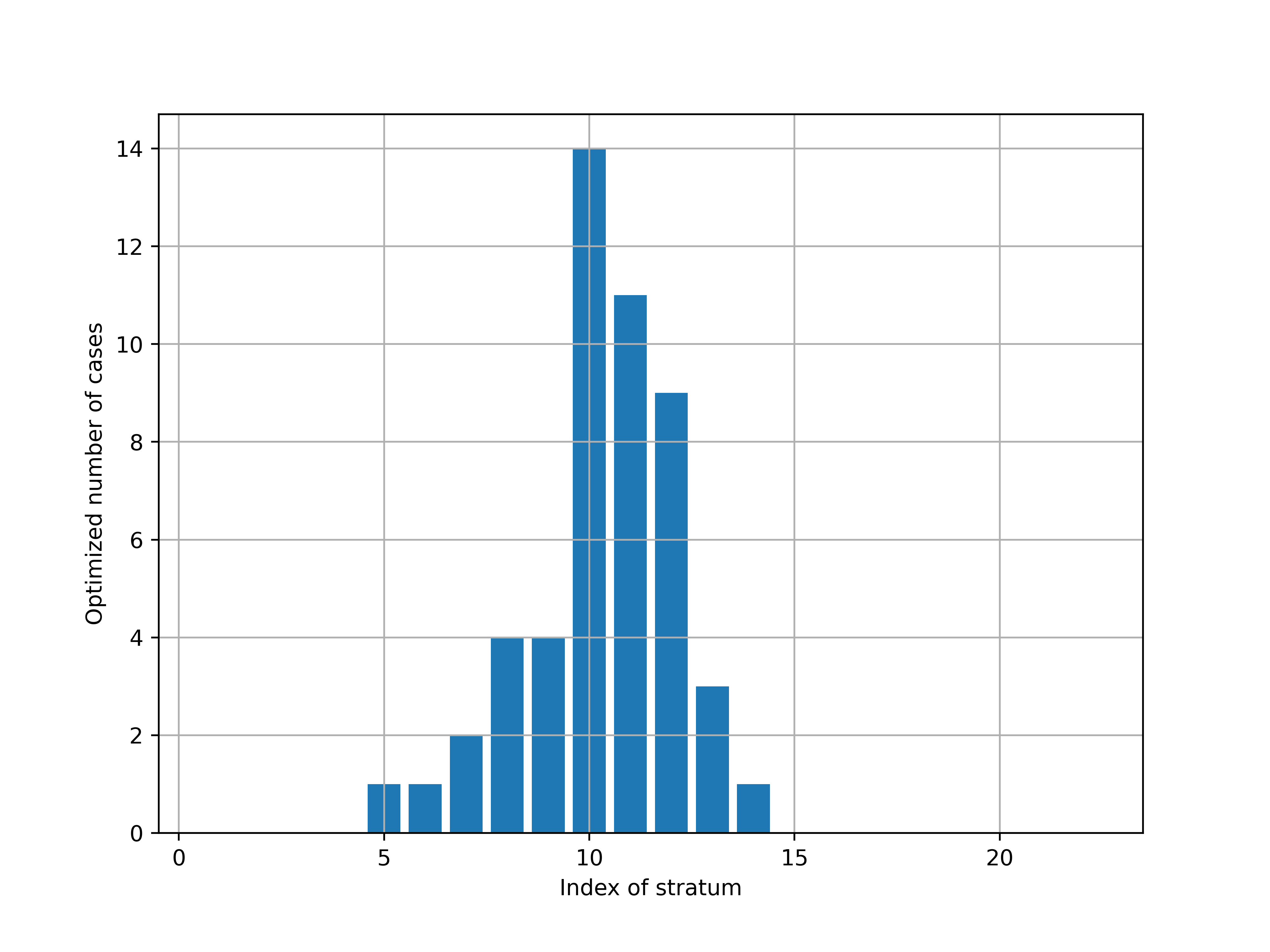}
  \caption{Left: Distribution of $P_i^{(1)} \sqrt{P_i^{(2)}\big(1 - P_i^{(2)}\big)}$.
           Right: Distribution of optimized number of cases.}
  \label{fig_eco_Ni_hist_iter2_1}
\end{figure}

\begin{figure}[h!]
  \centering
  \includegraphics[height=6cm,trim=0cm 0.5cm 2cm 1.5cm,clip]{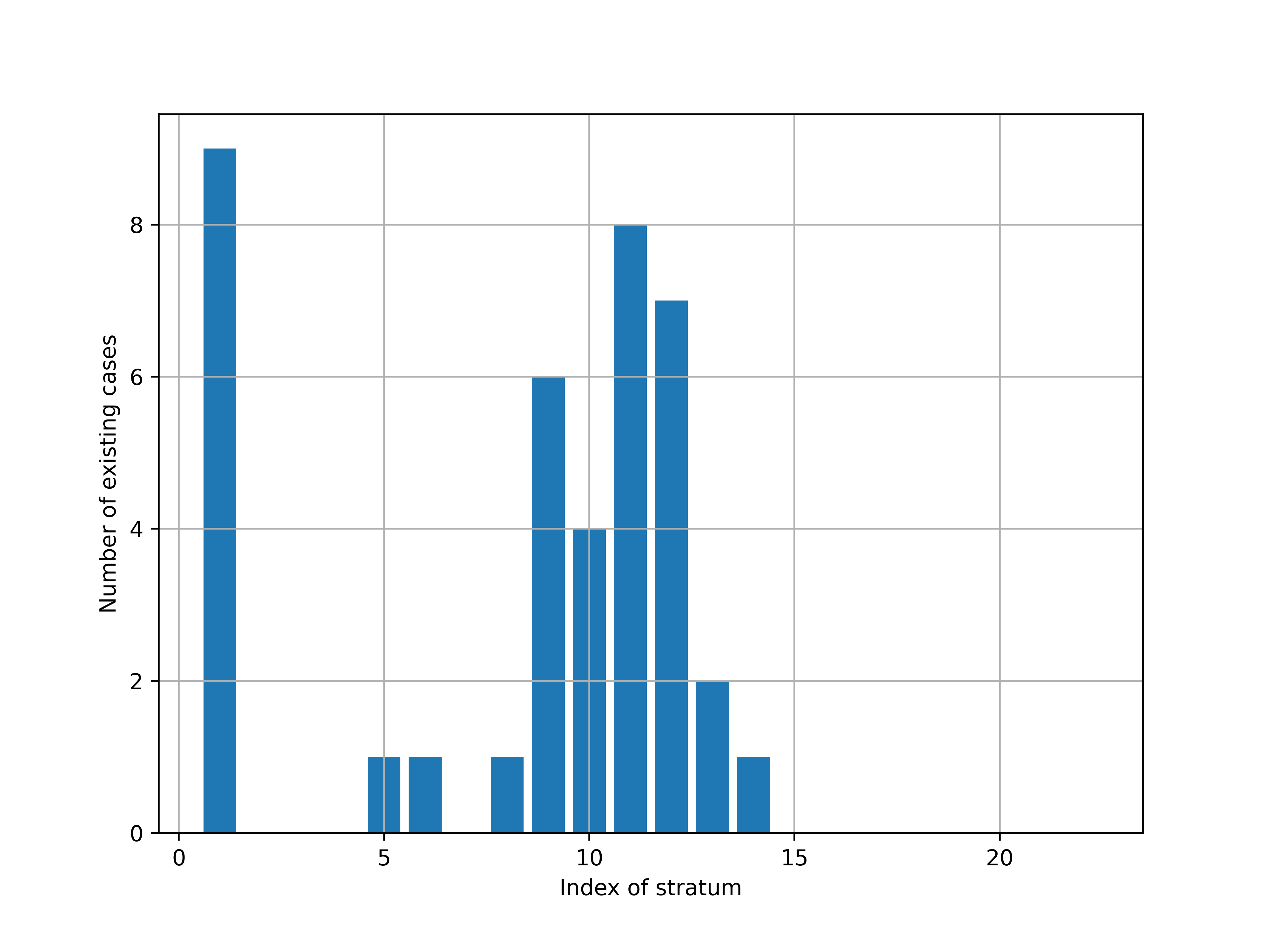}
  \includegraphics[height=6cm,trim=1cm 0.5cm 2cm 1.5cm,clip]{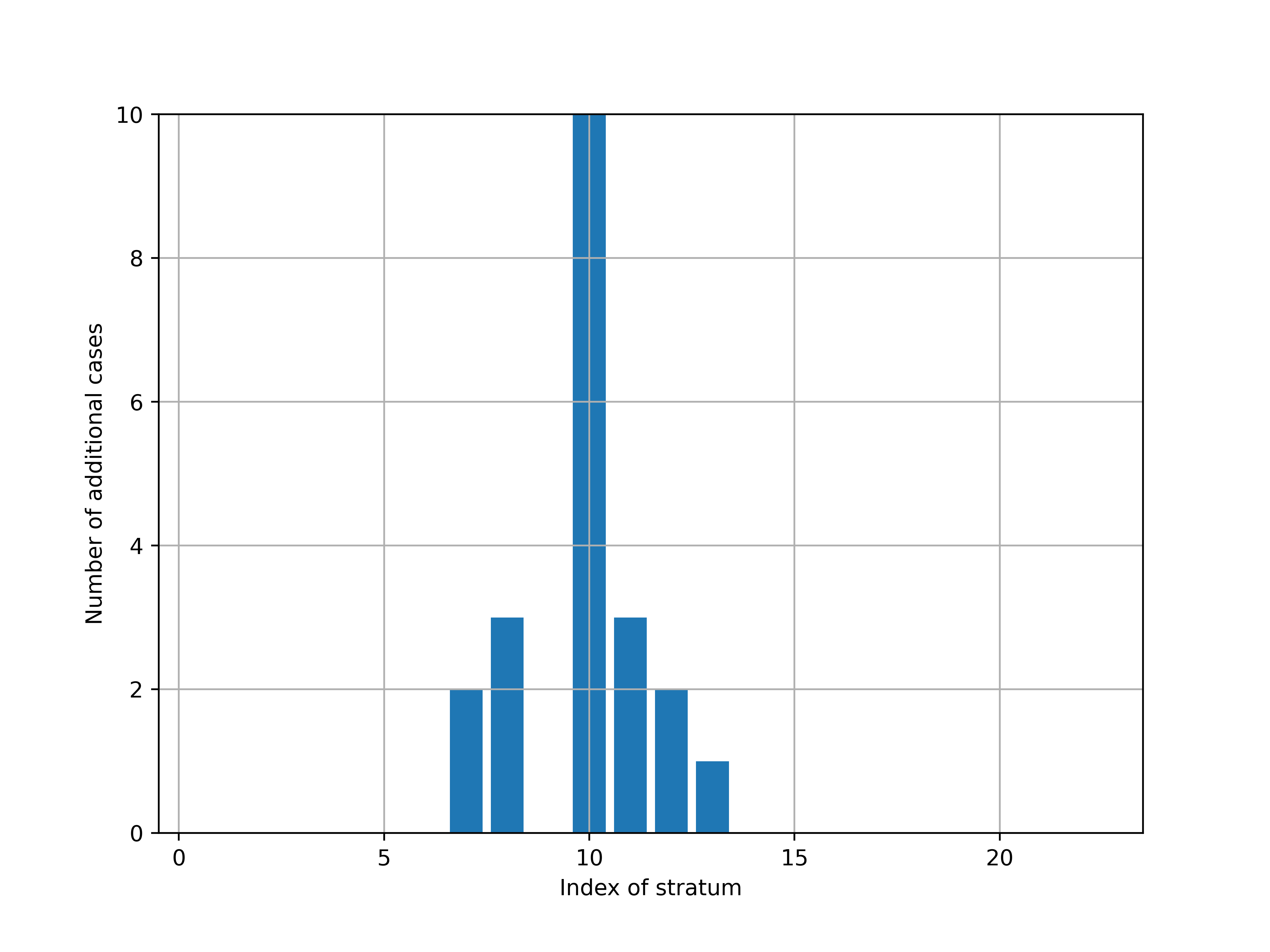}
  \caption{Left: Distribution of existing of cases.
           Right: Distribution of additional cases, where $\sum N_i = 21$.}
  \label{fig_eco_Ni_hist_iter2_2}
\end{figure}

The optimized distribution of $N_i$ for next adaptive iteration is shown in figure \ref{fig_eco_Ni_hist_iter2_1} and figure \ref{fig_eco_Ni_hist_iter2_2}.
Note that when calculating the number of additional cases for each stratum, we need to exclude the number of existing cases in that stratum.

\FloatBarrier

\FloatBarrier

\begin{figure}[h!]
  \centering
  \includegraphics[height=7cm,trim=0cm 0.5cm 0cm 1.5cm,clip]{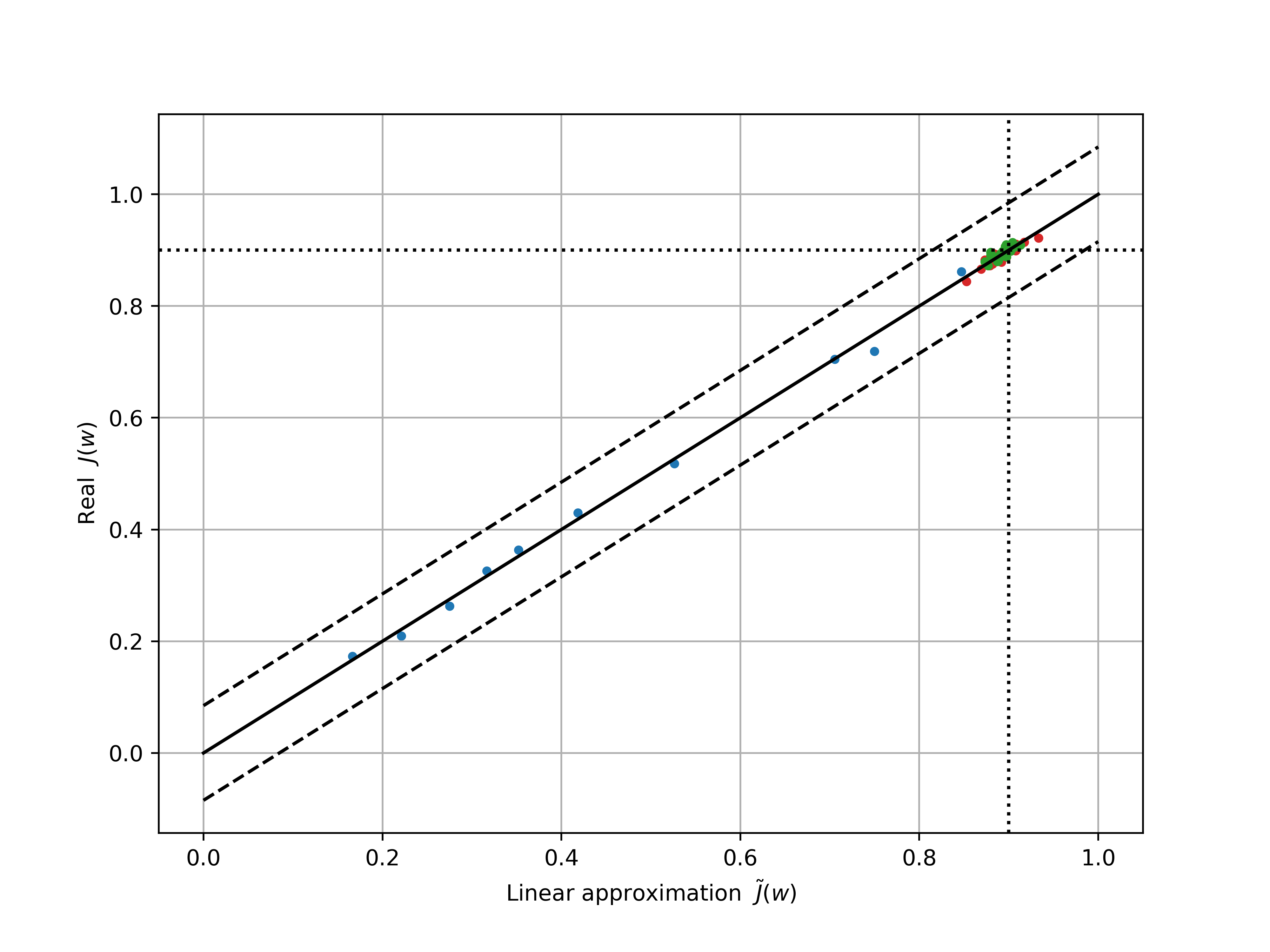}
  \caption{Real objective function $\MJ(w)$ versus the linear approximation $\tilde{\MJ}(w)$. The blue, red and green dots show the preliminary, first and second adaptive iteration, respectively.}
  \label{fig_eco_real_vs_approx_iter2_mixed}
\end{figure}

According to the $N_i$ distribution in figure \ref{fig_eco_Ni_hist_iter2_2}, we further run 21 additional cases in the second adaptive sampling iteration.
Utilizing the 21 additional flow solutions, we once again update the regression model, and further plot the real lift coefficients against the approximated lift coefficients in figure \ref{fig_eco_real_vs_approx_iter2_mixed}.

\FloatBarrier

\begin{figure}[h!]
  \centering
  \includegraphics[height=7cm,trim=1cm 0.5cm 2cm 1.5cm,clip]{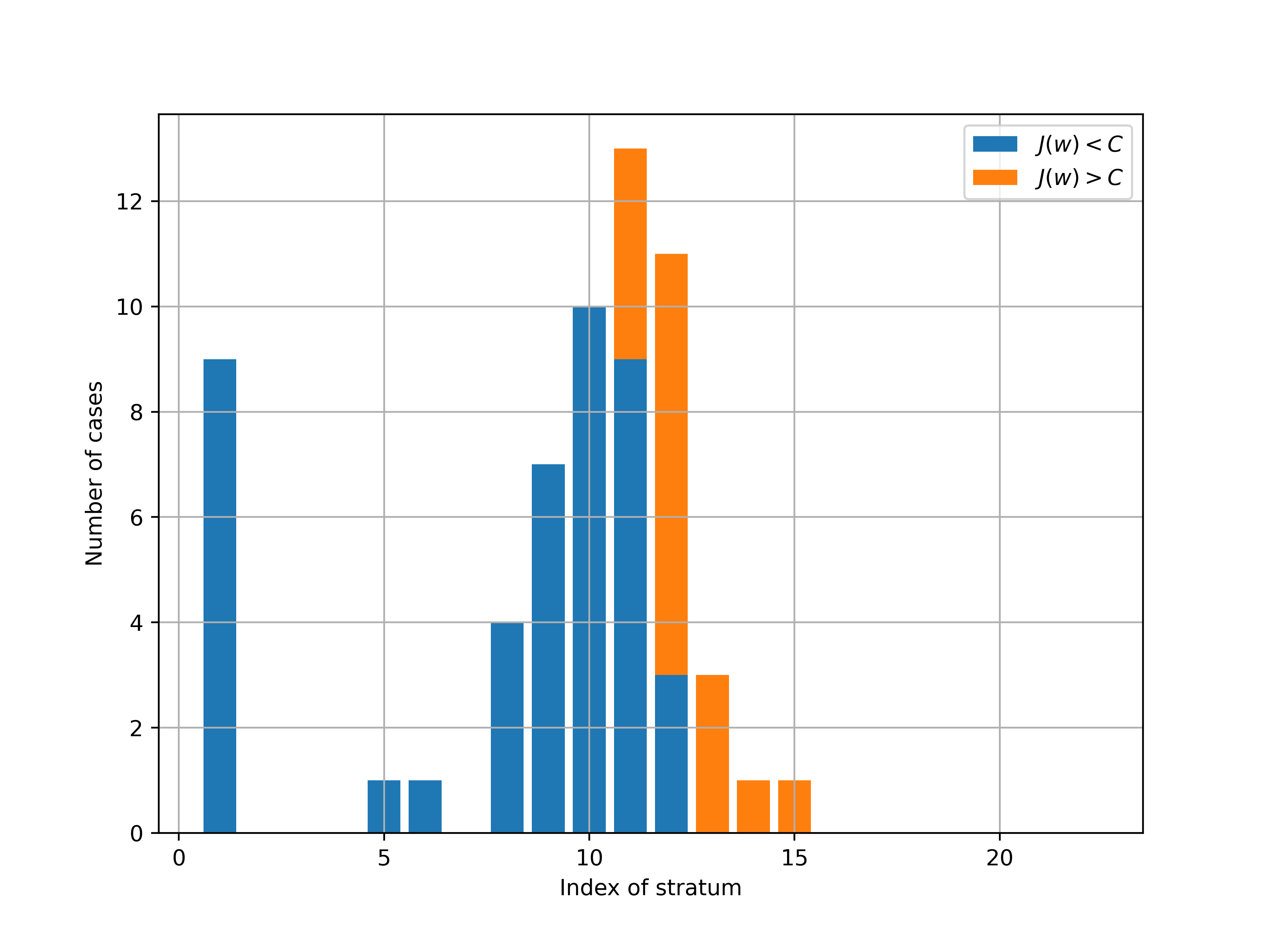}
  \caption{Number of cases distributed in different strata.
  Blue bars: number of cases in $S_i$ with $\MJ(w)<\MC$.
  Orange bars: number of cases in $S_i$ with $\MJ(w)>\MC$.}
  \label{fig_eco_NNC_hist_iter2_mixed}
\end{figure}
\FloatBarrier

\begin{figure}[h!]
  \centering
  \includegraphics[height=6cm,trim=0.5cm 0.5cm 2cm 1.5cm,clip]{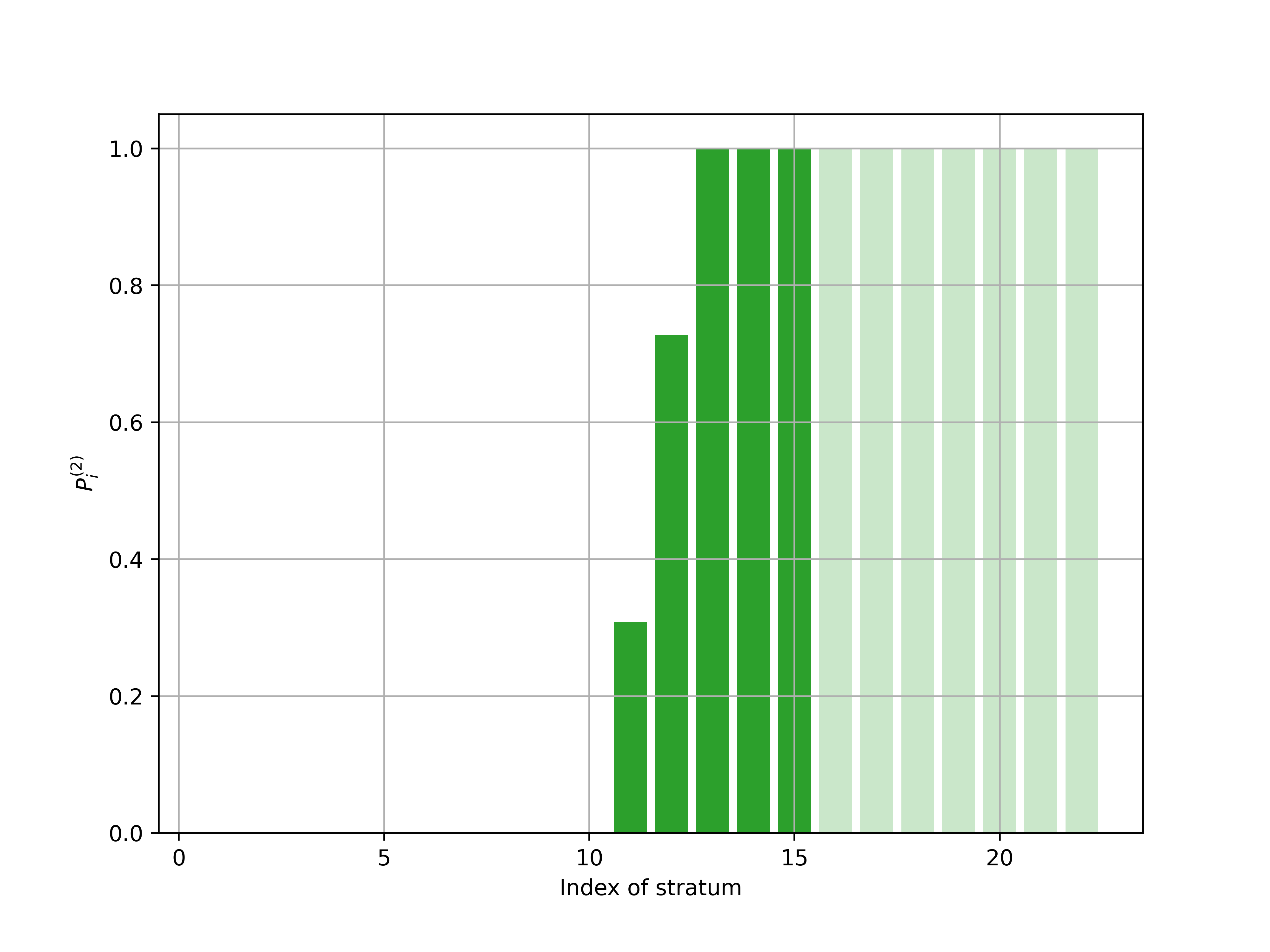}
  \includegraphics[height=6cm,trim=0.5cm 0.5cm 2cm 1.5cm,clip]{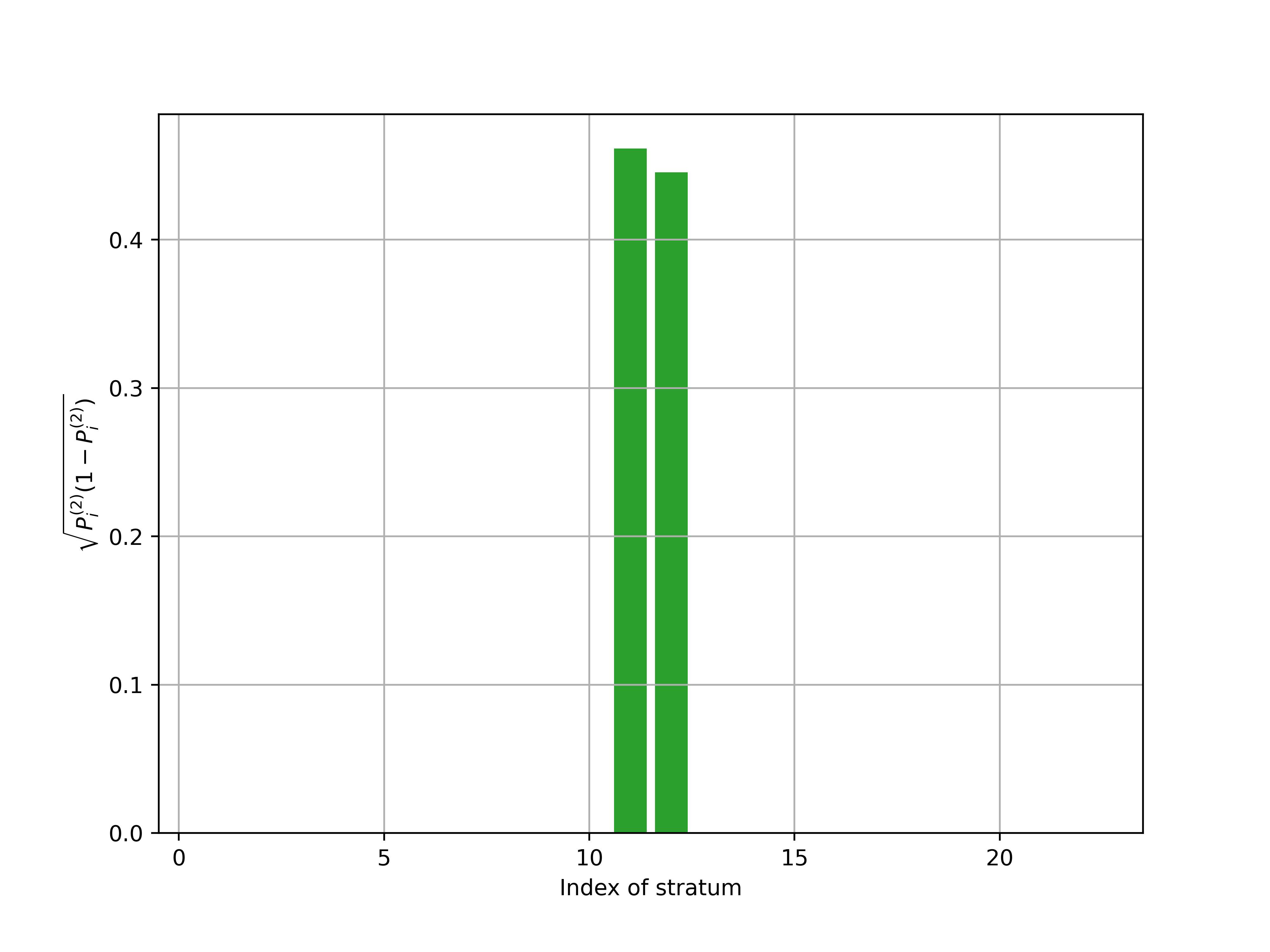}
  \caption{Left: Distribution of $P_i^{(2)}$, the solid bars are real flow solutions while the shaded bars are extrapolation. Right: Distribution of $\sqrt{P_i^{(2)}\big(1-P_i^{(2)}\big)}$.}
  \label{fig_eco_P2_hist_iter2}
\end{figure}
\FloatBarrier

After the second adaptive iteration,
\begin{equation}
\MP(\MJ(w) > \MC) \approx 0.00220
\end{equation}

The biased sample variance is,
\begin{equation}
\Var \bigg[\sum_{i=1}^{N_S} P_i^{(1)} P_i^{(2)}\bigg]
\approx \text{3.165626e-08}
\end{equation}

The unbiased population variance is,
\begin{equation}
s^2 \approx \text{3.449311e-08}
\end{equation}

Since we have $\mu = 0.00220, s = 0.000186$, the 95\% confidence interval is
\begin{equation}
(\mu-2s, \mu+2s) = (0.00183,0.00257) 
\end{equation}

\begin{table}[h!]
\centering
\begin{tabular}{c|c c c c c}
\hline
\hline
Iteration & $\MP(\MJ>\MC)$ & Biased variance & Unbiased variance & 95\% confidence interval & $N$\\
\hline
0         &               &                 &                   &                          & 100 \\
1         & 0.00213       & 5.191024e-08    & 6.847554e-08      & (0.00160, 0.00265)       & 99  \\
Total     &               &                 &                   &                          & 199 \\
\hline
0         &               &                 &                   &                          & 10 \\
1         & 0.00198       & 8.588410e-08    & 1.110937e-07      & (0.00131, 0.00265)       & 30 \\
2         & 0.00220       & 3.165626e-08    & 3.449311e-08      & (0.00183, 0.00257)       & 21 \\
Total     &               &                 &                   &                          & 61 \\
\hline
\hline
\end{tabular}
\caption{Advantage of multiple iterations with smaller $N$ per iteration. Top: single adaptive iteration. Bottom: multiple adaptive iterations.}
\label{tab_ensemble_size}
\end{table}
\FloatBarrier

\section{Conclusion}
This article presents an adaptive sampling approach for accurately estimating the probability of a rare event.
As an example, the adaptive sampling approach was implemented to estimate the probability of exceeding a critical lift coefficient for a set of parameterized geometries,
where the uncertainty was prescribed by 6 stochastic geometric and freestream parameters.
100 preliminary cases were first simulated.
Based on the solutions of these preliminary cases, a linear regression model was built and further applied to split the stochastic parameters into multiple strata.
The sampling space of stochastic geometric and freestream parameters was divided into 102 different strata, with 100 strata clustered around $\tilde{\MJ}(w) = 0.9$.
The adaptive sampling approach provided the optimized distribution of additional cases, such that the variance of the estimator was minimized.
Based on the lift coefficients calculated from 100 preliminary and 99 additional flow solutions, the probability of achieving a high lift coefficient was accurately estimated.
It has been shown that the adaptive sampling approach is hundreds of times more efficient than the brute-force Monte Carlo method, and the performance could be further improved by running multiple adaptive sampling iterations with less sampling points per iteration.

\bibliography{reference}

\begin{thebibliography}{8}
\newcommand{\enquote}[1]{``#1''}
\providecommand{\natexlab}[1]{#1}
\providecommand{\url}[1]{\texttt{#1}}
\providecommand{\urlprefix}{URL }
\expandafter\ifx\csname urlstyle\endcsname\relax
  \providecommand{\doi}[1]{\discretionary{}{}{}https://doi.org/#1}\else
  \providecommand{\doi}[1]{\discretionary{}{}{}\urlstyle{rm}\url{https://doi.org/#1}}\fi

\bibitem[{Iaccarino et~al.(2011)Iaccarino, Pecnik, Glimm, and
  Sharp}]{iaccarino2011qmu}
Iaccarino, G., Pecnik, R., Glimm, J., and Sharp, D., \enquote{A QMU approach
  for characterizing the operability limits of air-breathing hypersonic
  vehicles,} \emph{Reliability Engineering \& System Safety}, Vol.~96, No.~9,
  2011, pp. 1150--1160.

\bibitem[{Qiqi~Wang and Iaccarino(March 2012)}]{wang2012scramjet}
Qiqi~Wang, J. J.~A., Karthik~Duraisamy, and Iaccarino, G., \enquote{Risk
  Assessment of Scramjet Unstart Using Adjoint-Based Sampling Methods,}
  \emph{AIAA Journal}, Vol.~50, No.~3, March 2012.
\newblock \doi{10.2514/1.J051264}.

\bibitem[{Eldred et~al.(2002)Eldred, Giunta, Wojtkiewicz, and
  Trucano}]{eldred2002formulations}
Eldred, M., Giunta, A., Wojtkiewicz, S., and Trucano, T., \enquote{Formulations
  for surrogate-based optimization under uncertainty,} \emph{9th AIAA/ISSMO
  symposium on multidisciplinary analysis and optimization}, 2002, p. 5585.

\bibitem[{Giunta et~al.(2004)Giunta, Eldred, Swiler, Trucano, and
  Wojtkiewicz}]{giunta2004perspectives}
Giunta, A., Eldred, M., Swiler, L., Trucano, T., and Wojtkiewicz, S.,
  \enquote{Perspectives in Optimization Under Uncertainty: Algorithms and
  Applications,} \emph{10th AIAA/ISSMO Multidisciplinary Analysis and
  Optimization Conference}, 2004, p. 4451.

\bibitem[{Haimes and Dannenhoffer(2013)}]{haimes2013esp}
Haimes, R., and Dannenhoffer, J., \enquote{The engineering sketch pad: A
  solid-modeling, feature-based, web-enabled system for building parametric
  geometry,} \emph{21st AIAA Computational Fluid Dynamics Conference}, 2013, p.
  3073.

\bibitem[{Dannenhoffer and Haimes(2016)}]{dannenhoffer2016esp}
Dannenhoffer, J., and Haimes, R., \enquote{Generation of Multi-fidelity,
  Multi-discipline Air Vehicle Models with the Engineering Sketch Pad,}
  \emph{54th AIAA Aerospace Sciences Meeting}, 2016, p. 1925.

\bibitem[{Haimes(April 27, 2017)}]{haimes2017esp}
Haimes, B., \enquote{The Engineering Sketch Pad (ESP): Supporting Design
  Through Analysis,} \emph{Advanced Modeling \& Simulation (AMS) Seminar
  Series}, April 27, 2017.
\newblock
  \urlprefix\url{https://www.nas.nasa.gov/publications/ams/2017/04-27-17.html#:~:text=The\%20Engineering\%20Sketch\%20Pad\%20is,license\%20and\%20freely\%20available\%20here.}

\bibitem[{Pointwise(Retrieved November 10, 2020)}]{Pointwise}
Pointwise, I., \enquote{Glyph, Version 3.18.4,} , Retrieved November 10, 2020.
\newblock
  \urlprefix\url{https://www.pointwise.com/glyph2/files/Glyph/cxx/GgGlyph-cxx.html}.

\end{thebibliography}

\end{document}